\documentclass{article}
\usepackage{fullpage}
\usepackage{mathptmx}
\usepackage{amsmath, amsthm}
\usepackage{amssymb}
\usepackage{appendix}
\usepackage{color}
\usepackage{adjustbox}
\usepackage{listings}
\usepackage{wrapfig}
\usepackage{url}
\usepackage{tikz}
\usepackage{natbib}
\setcitestyle{authoryear,square}
\usetikzlibrary{positioning}

\newcommand{\comment}[1]{{}}

\newtheorem{Lem}{Lemma}

\newtheorem{Pro}[Lem]{Proposition}

\newtheorem{Def}{Definition}

\newtheorem{Ex}{Example}

\newcommand{\lsem}{\mbox{$[\![$}}
\newcommand{\rsem}{\mbox{$]\!]$}}
\newcommand{\sem}[1]{\mbox{$\lsem #1 \rsem$}}

\newcommand{\id}[1]{\mbox{\it #1\/}}

\def\p@enumiii{\theenumi(\theenumii)}

\newcommand\annotate[1]%
{\textcolor{blue}{$^\maltese$}%
\-\marginpar[\raggedleft\scriptsize \textcolor{red}{#1}]%
{\raggedright\scriptsize \textcolor{red}{#1}}}

\newcommand{\myparagraph}[1]{\paragraph{\textbf{#1}}}

\let\cite\citep
\newtheorem*{Pro1}{Proposition}

\title
{Constraint-Based Inference in
  Probabilistic~Logic~Programs}
\author{Arun Nampally,  Timothy Zhang, C.\ R.\ Ramakrishnan\\
Department of Computer Science, 
Stony Brook University, 
Stony Brook, NY 11794 \\
\url{{anampally, thzhang, cram}@cs.stonybrook.edu}
}

\begin{document}
\maketitle

\tikzset{vertex/.style={draw, rectangle, rounded corners}}
\tikzset{leaf/.style={draw, rectangle, rounded corners}}
\tikzset{dir_edge/.style={draw, thick, ->}}
\tikzset{ellipsis/.style={draw, dotted, ->}}
\tikzset{osddconstr/.style={draw, rectangle, rounded corners, red,
    fill=white, near start}}
\tikzset{osddconstr1/.style={draw, rectangle, rounded corners, red,
    fill=white, near end}}

\lstset{%
  language=Prolog,
  basicstyle=\ttfamily,
  commentstyle=\rmfamily\it\color{blue},
  columns=fullflexible,
}

\begin{abstract}
  Probabilistic Logic Programs (PLPs) generalize traditional logic programs and allow the encoding of models combining logical structure and uncertainty. In PLP, inference is performed by summarizing the possible worlds which entail the query in a suitable data structure, and using it to compute the answer probability.  Systems such as ProbLog, PITA, etc., use propositional data structures like explanation graphs, BDDs, SDDs, etc., to represent the possible worlds.  While this approach saves inference time due to substructure sharing, there are a number of problems where a more compact data structure is possible.  We propose a data structure called Ordered Symbolic Derivation Diagram (OSDD) which captures the possible worlds by means of constraint formulas.  We describe a program transformation technique to construct OSDDs via query evaluation, and give procedures to perform exact and approximate inference over OSDDs.  Our approach has two key properties.  Firstly, the exact inference procedure is a generalization of traditional inference, and results in speedup over the latter in certain settings.  Secondly, the approximate technique is a generalization of likelihood weighting in Bayesian Networks, and allows us to perform sampling-based inference with lower rejection rate and variance. We evaluate the effectiveness of the proposed techniques through experiments on several problems.
This paper is under consideration for acceptance in TPLP.
\end{abstract}
\section{Introduction}
\label{sec:intro}

A wide variety of models that combine logical and statistical
knowledge can be expressed succinctly in the Probabilistic Logic
Programming (PLP) paradigm. The expressive power of PLP goes beyond
that of traditional probabilistic graphical models (eg. Bayesian
Networks (BNs) and Markov Networks (MNs)) as can be seen in the
examples in Figs. \ref{fig:intro-ex-palindrome} and
\ref{fig:intro-ex-birthday}. These examples are written in PRISM, a
pioneering PLP language \cite{sato1997prism}. While the example in
Fig. \ref{fig:intro-ex-palindrome} encodes the probability
distribution of a palindrome having a specific number of occurrences
of a given character, the example in Fig. \ref{fig:intro-ex-birthday}
encodes the probability that at least two persons in a given set
have the same birthday. Examples such as these and other models like
reachability over graphs with probabilistic links illustrate how
logical clauses can be used to specify models that go beyond what is
possible in traditional probabilistic graphical models.

\begin{figure}
  \begin{minipage}{0.45\textwidth}
    \begin{lstlisting}[name=PalindromeExample]
% Generate a list of N random variables.
genlist(0, []).
genlist(N, L) :-
    N > 0,
    msw(flip, N, X),
    L = [X|L1],
    N1 is N-1,
    genlist(N1, L1).
% Evidence: list is a palindrome.
evidence(N) :-
    genlist(N, L), palindrome(L).
% Query: string has K 'a's
query(N, K) :-
    genlist(N, L), count_as(L, K).
    \end{lstlisting}
  \end{minipage}
  \hfill
  \begin{minipage}{0.45\textwidth}
    \begin{lstlisting}[name=PalindromeExample]
% Check if a given list is a palindrome
palindrome(L) :- phrase(palindrome, L).
palindrome --> [].
palindrome --> [_X].
palindrome --> [X], palindrome, [X].
% Query condition:
count_as([], 0).
count_as([X|Xs], K) :-
    count_as(Xs, L),
    (X=a -> K is L+1; K=L).
% Domains:
values(flip, [a,b]).
% Distribution parameters:
set_sw(flip, [0.5, 0.5]).
    \end{lstlisting}
  \end{minipage}
  \caption{Palindrome PLP}
  \label{fig:intro-ex-palindrome}
\end{figure}

\paragraph{\textbf{The Driving Problem.}}
The expressiveness of PLP comes at a cost.  Since PLP is an extension
to traditional logic programming, inference in PLP is undecidable in
general.  Inference is intractable even under strong finiteness
assumptions.  For instance, consider the PRISM program in
Fig.~\ref{fig:intro-ex-palindrome}.  In that program, \texttt{genlist/2}
defines a list of the outcomes of $N$ identically distributed random
variables ranging over \texttt{\{a,b\}} (through \texttt{msw/3}
predicates).  Predicate \texttt{palindrome/1} tests, using a definite
clause grammar definition, if a given list is a palindrome; and
\texttt{count\_as/2} tests if a given list contains $k$ (not
necessarily consecutive) ``\texttt{a}''s.  Using these predicates,
consider the inference of the conditional probability of
$\mathtt{query}(n,k)$ given $\mathtt{evidence}(n)$: i.e., the
probability that an $n$-element palindrome has $k$ ``\texttt{a}''s.

\begin{wrapfigure}{R}{0.4\textwidth}
  \vspace{-25pt}
  \centering
\begin{lstlisting}[name=BirthdayExample]
% Two from a population of
% size N share a birthday.
same_birthday(N) :-
    person(N, P1),
    msw(b, P1, D), 
    person(N, P2),
    P1 < P2,
    msw(b, P2, D)
% Bind P, backtracking
% through 1..N    
person(N, P) :-
    basics:for(P, 1, N).
% Distribution parameters:
:- set_sw(b(_),
    uniform(1,365)).
\end{lstlisting}
\caption{Birthday PLP}
\label{fig:intro-ex-birthday}
\end{wrapfigure}

The conditional probability is well-defined according to PRISM's
distribution semantics~\cite{sato2001parameter}.  However, \emph{
  PRISM itself will be unable to correctly compute the conditional
  query's probability}, since the conditional query, as encoded above,
will violate the PRISM system's assumptions of independence among
random variables used in an explanation.  Moreover, while the
probability of goal \verb|evidence(N)| can be computed in linear time
(using explanation graphs), the conditional query itself is
intractable, since the computation is dominated by the binomial
coefficient $\binom{N}{k}$.  This is not surprising since probabilistic
inference is intractable over even simple statistical models such as
Bayesian networks.  Consequently, exact inference techniques used in
PLP systems such as PRISM, ProbLog~\cite{de2007problog} and
PITA~\cite{riguzzi2011pita}, 
have exponential time complexity when used on such programs.

Approximate inference based on rejection sampling also performs poorly,
rejecting a vast number of generated samples, since the likelihood of
a string being a palindrome decreases exponentially in $N$.
Alternatives such as Metropolis-Hastings-based Markov Chain Monte
Carlo (MCMC) techniques~\cite[e.g.]{hastings1970monte} do not behave
much better:   the chains exhibit poor convergence
(mixing), since most transitions lead to strings inconsistent with
evidence.  Gibbs-sampling-based MCMC~\cite{geman1984stochastic} cannot
be readily applied since the dependencies between random variables are
hidden in the program and not explicit in the model.

\paragraph{\textbf{Our Approach.}}  

In this paper, we use PRISM's syntax and distribution semantics,
\emph{but without the requirements imposed by the PRISM system}, namely, that distinct explanations of an answer are pairwise mutually exclusive and all random variables within an explanation are independent.  We, however, retain the assumption that distinct random variable instances are independent.   Thus we consider PRISM programs with
their \emph{intended} model-theoretic semantics, rather than that
computed by the PRISM system.
\label{para:prism-assumptions}

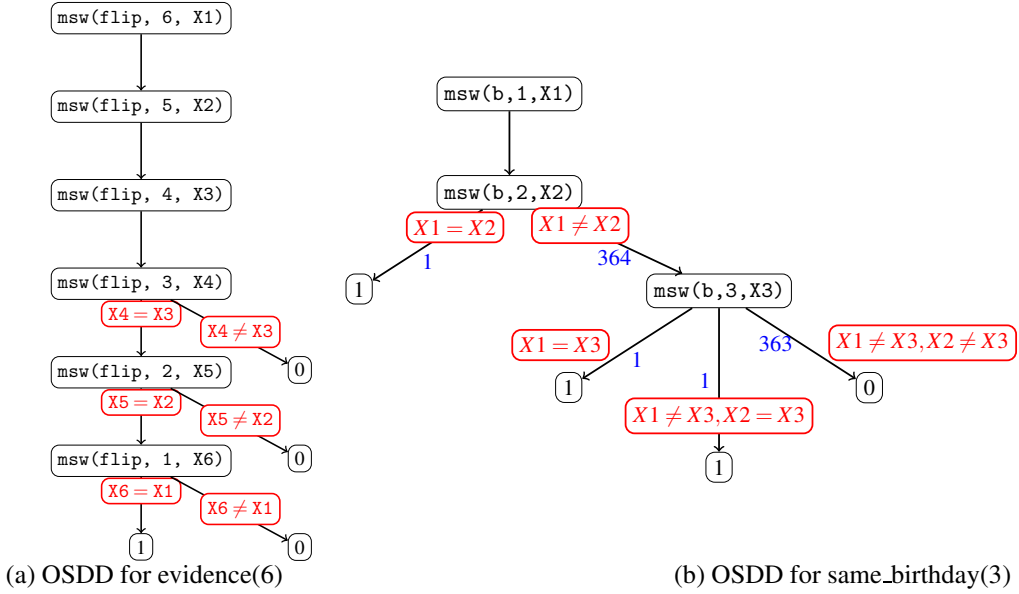
\begin{figure}[h]
  \centering
  \begin{tabular}{cp{0.0in}c}
    \begin{minipage}[t]{0.15\textwidth}
  \begin{adjustbox}{scale=0.75}
    $\vcenter{
      \begin{tikzpicture}
      \node[vertex] (a) {\texttt{msw(flip, 6, X1)}}; 
      \node[vertex] (b) [below=of a] {\texttt{msw(flip, 5, X2)}}; 
      \node[vertex] (c) [below=of b] {\texttt{msw(flip, 4, X3)}}; 
      \node[vertex] (d) [below=of c] {\texttt{msw(flip, 3, X4)}};
      \node[leaf] (fail1) [below right=of d] {$0$};
      \node[vertex] (e) [below=of d] {\texttt{msw(flip, 2, X5)}};
      \node[leaf] (fail2) [below right=of e] {$0$};
      \node[vertex] (f) [below=of e] {\texttt{msw(flip, 1, X6)}};
      \node[leaf] (fail3) [below right=of f] {$0$};
      \node[leaf] (g) [below=of f] {$1$};
      \path[dir_edge] (a) -- (b); 
      \path[dir_edge] (b) -- (c);
      \path[dir_edge] (c) -- (d);
      \path[dir_edge] (d) -- node[osddconstr] {$\mathtt{X4}=\mathtt{X3}$} (e);
      \path[dir_edge] (e) -- node[osddconstr] {$\mathtt{X5}=\mathtt{X2}$} (f);
      \path[dir_edge] (f) -- node[osddconstr] {$\mathtt{X6}=\mathtt{X1}$} (g);
      \path[dir_edge] (d) -- node[osddconstr, below right] {$\mathtt{X4} \neq \mathtt{X3}$} (fail1);
      \path[dir_edge] (e) -- node[osddconstr, below right] {$\mathtt{X5} \neq \mathtt{X2}$} (fail2);
      \path[dir_edge] (f) -- node[osddconstr, below right] {$\mathtt{X6} \neq \mathtt{X1}$} (fail3);
    \end{tikzpicture}
  }$
  \end{adjustbox}
\end{minipage}
& &
\begin{minipage}{0.8\textwidth}
  \begin{adjustbox}{scale=0.85}
  	$\vcenter{
    \begin{tikzpicture}
      \node[vertex] (a) {\texttt{msw(b,1,X1)}}; 
      \node[vertex] (b) [below=of a] {\texttt{msw(b,2,X2)}};
      \node[leaf] (s1) [below left=of b] {$1$};       
      \node[vertex] (c) [below right=of b] {\texttt{msw(b,3,X3)}};
      \node (dummy) [below=of c] {};
      \node[leaf] (s2) [below left=of c] {$1$}; 
      \node[leaf] (s3) [below=of dummy] {$1$}; 
      \node[leaf] (f1) [below right=of c] {$0$}; 

      \path[dir_edge] (a) -- (b); 
      \path[dir_edge] (b) -- node[osddconstr] {$X1=X2$} node[below, blue] {1} (s1);
      \path[dir_edge] (b) -- node[osddconstr] {$X1 \neq X2 $} node[below, blue] {364} (c);
      \path[dir_edge] (c) -- node[osddconstr1, above left] {$X1=X3$} node[below, blue] {1} (s2);
      \path[dir_edge] (c) -- node[osddconstr1] {$X1 \neq X3, X2=X3$} node[left,blue] {1} (s3);
      \path[dir_edge] (c) -- node[osddconstr1, above right]
      {$X1 \neq X3, X2 \neq X3$} node[left,blue] {363} (f1);
    \end{tikzpicture}
}$
  \end{adjustbox}
\end{minipage}\\
    (a) OSDD for evidence(6) && (b) OSDD for same\_birthday(3)
  \end{tabular}
  \caption{OSDDs for introductory examples}
  \label{fig:osdd-intro-ex}
\end{figure}

We propose a data structure called \emph{Ordered Symbolic Derivation
  Diagram} (OSDD) which represents the set of possible explanations
for a goal symbolically. The key characteristic of OSDDs is the use of
constraints on random variables. This data structure is constructed
through tabled evaluation on a transformed input program. For example,
the OSDD for ``same\_birthday(3)'' from example
Fig. \ref{fig:intro-ex-birthday} is shown in
Fig. \ref{fig:osdd-intro-ex}(b). This data structure can be used for
performing exact inference in polynomial time as will be described
later. In cases where exact inference is intractable, OSDDs can be
used to perform sampling based inference. For example, the OSDD for
``evidence(6)'' from example Fig. \ref{fig:intro-ex-palindrome} is shown in
Fig. \ref{fig:osdd-intro-ex} (a) and can be used for performing
likelihood weighted sampling
\cite{fung1990weighting,shachter1990simulation}.

The rest of the paper is organized as follows. In Section
\ref{sec:osdd} we formally define OSDDs and the operations on
OSDDs. Next we give the procedure for construction of OSDDs. This
procedure relies on a program transformation which is explained in
Section \ref{sec:construction}. Next we give the exact and approximate
inference algorithms using OSDDs in Section
\ref{sec:inference}.  We present the experimental results in
Section \ref{sec:expt}, related work in Section \ref{sec:related}, and concluding remarks in Section~\ref{sec:conclusion}.

\section{Ordered Symbolic Derivation Diagrams}
\label{sec:osdd}
\myparagraph{Notation:} We assume familiarity with common logic
programming terminology such as variables, terms, substitutions,
predicates and clauses. We use Prolog's convention, using identifiers
beginning with a lower case letter to denote atomic constants or
function symbols, and those beginning with an upper case letter to
denote variables.  We often use specific symbols such as $t$ to denote
ground terms, and $i,j,k$ to denote integer indices.  We assume an
arbitrary but fixed ordering "$\prec$" among variables and ground
terms.

A \textbf{\emph{type}} is a finite collection of ground terms.  In
this paper, types represent the space of outcomes of switches or
random processes.  For example in Palindrome example of Fig
\ref{fig:intro-ex-palindrome}, the set of values $\{a,b\}$ is a type.
A variable $Y$ referring to the outcome of a switch is a typed
variable; its type, denoted $type(Y)$, is deemed to be the same as the
space of outcomes of $s$. The type of a ground term $t$ can be any of
the sets it is an element of.

\begin{Def}[Atomic Constraint]
  An atomic constraint, denoted $\beta$, is of the form $\{X=T\}$ or
  $\{X \neq T\}$, where $X$ is a variable and $T$ is a variable or a
  ground term of the same type as $X$.  When $T$ is a ground term, we
  can assert that $type(T) = type(X)$.
\end{Def}

A set of atomic constraints representing their conjunction is called a
\textbf{\emph{constraint formula}}. Constraint formulas are denoted by
symbols $\gamma$ and $\phi$. Note that atomic constraints are closed
under negation, while  constraint formulas are not closed
under negation.

\begin{Def}[Constraint Graph]
  \label{def:constraint-graph}
  The constraint graph for a constraint formula $\gamma$ is labeled
  undirected graph, whose nodes are variables and ground terms in
  $\gamma$.  Edges are labeled with "$=$" and "$\not=$" such that
  \begin{itemize}
  \item $T_1$ and $T_2$ are connected by an edge labeled "$=$" if, and
    only if, $T_1 = T_2$ is entailed by $\gamma$.
  \item $T_1$ and $T_2$ are connected by an edge labeled "$\not=$" if,
    and only if, $T_1 \not = T_2$ is entailed by $\gamma$, and at
    least one of $T_1, T_2$ is a variable.
   \end{itemize}
\end{Def}
Note that a constraint graph may have edges between two terms even
when there is no explicit constraint on the two terms in $\gamma$.

\begin{Def}[Ordering]
  \label{def:ordering}
  Based on an (arbitrary but fixed) ordering among ground terms, we
  define an ordering on ground switch instance pairs $(s,i)$ and
  $(s',i')$ as follows:
  \begin{itemize}
  \item If $i \prec i'$ then
    $(s,i) \prec (s',i')$ for all ground terms $s$ and  $s'$.
  \item If $i=i' \land s \prec s'$ then
    $(s,i) \prec (s',i')$.
  \item If $i=i' \land s=s'$ then $(s,i)=(s',i')$.
  \end{itemize}
\end{Def}

\myparagraph{Canonical representation of constraint formulas.}  A
constraint formula is represented by its constraint graph which in
turn can be represented as a set of triples
$(source, destination, constraint)$ each of which represents an edge
in the constraint graph.  Recall that variables and ground terms can
be compared using total order $\prec$.  Assuming an order between the
two symbols "$=$" and "$\not=$", we can define a lexicographic order
over each edge triple, and consequently order the elements of the edge
set.  The sequence of edge triples following the above order is a
canonical representation of a constraint formula.  Using this
representation, we can define a total order, called the
\emph{canonical order}, among constraint formulas themselves given by
the lexicographic ordering defined over the edge sequences.

Given any constraint formula
$\gamma = \{\beta_1, \beta_2, \ldots, \beta_n\}$, its negation
$\neg \gamma$ is given by a set of constraint formulas
$\neg \gamma = \{\{\neg \beta_1\}, \{\beta_1,\neg \beta_2\}, \ldots,
\{\beta_1, \beta_2, \ldots, \beta_{n-1}, \neg \beta_n\}\}$.  The above
defines negation of a formula to be a set of constraint formulas which
are pairwise mutually exclusive and together represent the
negation.

The set of solutions of a constraint formula $\gamma$ is denoted
$\sem{\gamma}$ and their projection onto a variable
$X \in Vars(\gamma)$ is denoted $\sem{\gamma}_X$. The constraint
formula is unsatisfiable if $\sem{\gamma}=\emptyset$, and satisfiable
otherwise.  Note that substitutions can also be viewed as constraint
formulas.


\comment{
\myparagraph{PRISM.}  We briefly recall the syntax and semantics of
PRISM programs (see Appendix A for a more detailed overview).  PRISM
programs have Prolog-like syntax.  In a PRISM program the \texttt{msw}
relation (``multi-valued switch'') has a special meaning:
\texttt{msw(X,I,V)} says that \texttt{V} is the outcome of the
\texttt{I}-th instance from a family \texttt{X} of random processes
also called as \emph{switches}.  The set of variables
$\{\mathtt{V}_i \mid \mathtt{msw(}p, i, \mathtt{V}_i\mathtt{)}\}$ are
i.i.d. for a given random process $p$.  Note that the ternary
\texttt{msw} is used in the definition of the PRISM
language~\cite{sato2001parameter} but the PRISM system uses a binary
\texttt{msw} by omitting the instance parameter.

The meaning of a PRISM program is given in terms of a
\emph{distribution semantics}~\cite{sato1997prism,sato2001parameter}.
Inference in the PRISM system computes the semantics for a subset of
PRISM programs as an extension of tabled evaluation.  When the goal
selected at a step is of the form \texttt{msw(X,I,Y)}, then \texttt{Y}
is bound to a possible outcome of a random process \texttt{X}.
\emph{Thus in PRISM, derivations are constructed by enumerating the
  possible outcomes of each random variable.}  The derivation step is
associated with the probability of this outcome.  Under independence
and mutual exclusion assumptions, the probabilities can be computed
without materializing the derivations.
}

\begin{wrapfigure}{r}{0.48\textwidth}
  \vspace{-5pt}
  \begin{center}
    \includegraphics[width=0.39\textwidth]{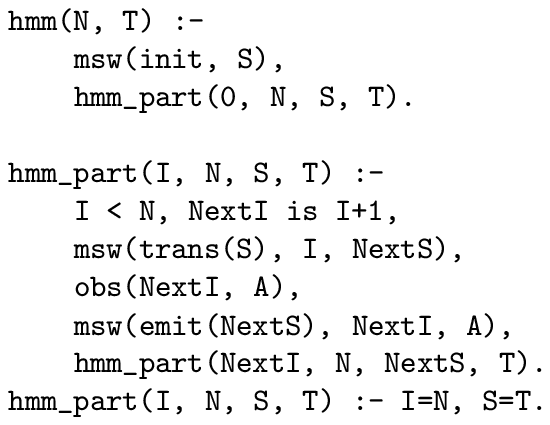}
  \end{center}
  \caption{PRISM program for an HMM}
  \label{fig:hmm-prism}
  \vspace{-5pt}
\end{wrapfigure}

\myparagraph{PRISM.}  The following is a high-level overview of PRISM.
PRISM programs have Prolog-like syntax (see Fig.~\ref{fig:hmm-prism}).
In a PRISM program the \texttt{msw} relation (``multi-valued switch'')
has a special meaning: \texttt{msw(X,I,V)} says that \texttt{V} is the
outcome of the \texttt{I}-th instance from a family \texttt{X} of
random processes. The set of variables
$\{\mathtt{V}_i \mid \mathtt{msw(}p, i, \mathtt{V}_i\mathtt{)}\}$ are
i.i.d. for a given random process $p$.

The distribution parameters of the random variables are specified
separately.

The program in Fig.~\ref{fig:hmm-prism} encodes a Hidden Markov Model
(HMM) in PRISM. 
The set of observations is encoded as facts of predicate \texttt{obs},
where \texttt{obs(I,V)} means that value \texttt{V} was observed at
time \texttt{I}.  In the figure, the clause defining \texttt{hmm} says
that \texttt{T} is the \texttt{N}-th state if we traverse the HMM
starting at an initial state \texttt{S} (itself the outcome of the
random process \texttt{init}).  In \texttt{hmm\_part(I, N, S, T)},
\texttt{S} is the \texttt{I}-th state, \texttt{T} is the \texttt{N}-th
state. The first clause of \texttt{hmm\_part} defines the conditions
under which we go from the \texttt{I}-th state \texttt{S} to the
\texttt{I+1}-th state \texttt{NextS}.  Random processes
\texttt{trans(S)} and \texttt{emit(S)} give the distributions of
transitions and emissions, respectively, from state \texttt{S}.

The meaning of a PRISM program is given in terms of a
\emph{distribution semantics}~\cite{sato1997prism,sato2001parameter}.
A PRISM program is treated as a non-probabilistic logic program over a
set of probabilistic facts, the \texttt{msw} relation.  An instance of
the \texttt{msw} relation defines one choice of values of all random
variables.  A PRISM program is associated with a set of least models,
one for each \texttt{msw} relation instance.  A probability
distribution is then defined over the set of models, based on the
probability distribution of the \texttt{msw} relation instances.  This
distribution is the semantics of a PRISM program.  Note that the
distribution semantics is declarative.  For a subclass of programs,
PRISM has an efficient procedure for computing this semantics based on
OLDT resolution~\cite{tamaki1986old}.
	
Inference in PRISM proceeds as follows.  When the goal selected at a
step is of the form \texttt{msw(X,I,Y)}, then \texttt{Y} is bound to a
possible outcome of a random process \texttt{X}.  \emph{Thus in PRISM,
  derivations are constructed by enumerating the possible outcomes of
  each random variable.}  The derivation step is associated with the
probability of this outcome.  If all random processes encountered in a
derivation are independent, then the probability of the derivation is
the product of probabilities of each step in the derivation.  If a set
of derivations are pairwise mutually exclusive, the probability of the
set is the sum of probabilities of each derivation in the set.
PRISM's evaluation procedure is defined only when the independence and
exclusiveness assumptions hold.  Finally, the probability of an answer
is the probability of the set of derivations of that answer.

\myparagraph{OSDD.} We use OSDDs to materialize derivations in a
compact manner.  OSDDs share a number of features with Binary Decision
Diagrams (BDDs)~\cite{BDD} and Multivalued Decision Diagrams
(MDDs)~\cite{MDDs}.  BDDs are directed acyclic graphs representing
propositional boolean formulas, with leaves labeled from $\{0,1\}$ and
internal nodes labeled with propositions.  In a BDD, each node has two
outgoing edges labeled $0$ and $1$, representing a true and false
valuation, respectively, for the node's proposition.  An MDD
generalizes a BDD where internal nodes are labeled with finite-domain
variables, and the outgoing edges are labeled with the possible
valuations of that variable.  In an OSDD, internal nodes represent
switches and the outgoing edges are labeled with constraints
representing the possible outcomes of that node's switch.

\begin{Def}[Ordered Symbolic Derivation Diagram]
  \label{def:osdd}
  An ordered symbolic derivation diagram over a set of typed variables
  $V$ is a tree, where leaves are labeled from the set $\{0,1\}$ and
  internal nodes are labeled by triples of the form $(s, k, Y)$, where
  $s$ and $k$ are switch and instance respectively and $Y \in V$. We
  call $Y$ the output variable of the node.  The edges are labeled by
  constraint formulas over $V$.  We represent OSDDs by textual
  patterns $(s,k,Y)[\gamma_i:\psi_i]$ where $(s,k,Y)$ is the label of
  the root and each sub-OSDD $\psi_i$ is connected to the root by an
  edge labeled $\gamma_i$.  OSDDs satisfy the following conditions:
  \begin{enumerate}
  \item{\textbf{Ordering:}} For internal nodes $n = (s,k,Y)$ and
    $n'= (s',k',Y')$, if $n$ is the parent of $n'$, then
    $(s,k)\prec(s',k')$. The edges are ordered by using the canonical
    ordering of the constraint formulas labeling them.
  \item{\textbf{Mutual Exclusion:}} The constraints labeling the
    outgoing edges from an internal node are pairwise mutually
    exclusive (i.e., for each $\gamma_i$ and $\gamma_j$,
    $\sem{\gamma_i \land \gamma_j} = \emptyset$).
  \item{\textbf{Completeness:}} Let $(s, k, Y)[\gamma_i : \psi_i]$ be
    a sub-OSDD and let $\sigma$ be any substitution that satisfies all
    constraints on the path from root to the given sub-OSDD such that
    $\sigma(Y) \in \id{type}(Y)$.  Then there is a $i$ such that
    $\sigma$ satisfies $\gamma_i$.
  \item{\textbf{Urgency:}} Let $\mathcal{O}(n)$ be the set of output
    variables in the path from the root to node $n$ (including
    $n$). Then for every constraint formula $\gamma_i$ labeling an
    outgoing edge from $n$, it holds that
    $Vars(\gamma_i) \subseteq \mathcal{O}(n)$ and for every ancestor
    $n'$ of $n$, $Vars(\gamma_i) \not \subseteq \mathcal{O}(n')$.
  \item{\textbf{Explicit constraints:}} If constraint formula
    $\gamma_i$ out of a node $n$ entails an implicit atomic constraint
    $\beta$ on variables in $\mathcal{O}(n)$, then $\beta$ occurs
    explicitly in the path from root to $n$.  A consequence of this
    condition is that the conjunction of constraint formulas labeling
    edges in a path will be satisfiable.
  \end{enumerate}
\end{Def}
A tree which satisfies all conditions of an OSDD \emph{except}
conditions 4 and 5 is called an \emph{improper OSDD}.

\begin{Ex}[OSDD properties]
  We illustrate the definiton by using the OSDD shown in
  Fig. \ref{fig:osdd-intro-ex}(b). The OSDD is represented by the
  textual pattern $(b,1,X1)[\emptyset : \psi_1]$ where $\psi_1$ is the
  sub-tree rooted at the node labeled $msw(b,2,X2)$, which in turn can
  be represented by the textual pattern
  $(b,2,X2)[\{X1=X2\}: 1, \{X1\neq X2\} : \psi_2]$ where $\psi_2$ is
  the sub-tree rooted at the node $msw(b,3,X3)$ and so on.  The
  internal nodes satisfy the total ordering based on the instance
  numbers. All outgoing edges from an internal node are pairwise
  mutually exclusive. For instance for the outgoing edges of the node
  labeled $msw(b, 3, X3)$, $X1=X3$ is mutually exclusive w.r.t to
  $X1\neq X3, X2=X3$ and $X1\neq X3, X2 \neq X3$. Similarly,
  $X1\neq X3, X2=X3$ is mutually exclusive to $X1\neq X3, X2 \neq
  X3$. Consider the sub-OSDD rooted at $msw(b, 2, X2)$ any
  substitution that grounds $X1, X2$ satisfies the (empty) constraints
  on the path from the root to that sub-tree. Further any such
  substitution will satisfy exactly one of the edge constraints
  $X1=X2$ or $X1\neq X2$. It is obvious from the example that urgency
  is satisfied. Finally consider the constraint formula
  $X1\neq X3, X2=X3$. This entails the implicit constraint
  $X1\neq X2$. However, this constraint is explicitly found in the
  path from the root to that sub-tree. Therefore the OSDD in
  Fig. \ref{fig:osdd-intro-ex}(b) is a proper OSDD.
\end{Ex}

OSDDs can be viewed as representing a set of explanations or
derivations where a node of the form $(s, k, Y)$ \emph{binds} $Y$.
This observation leads to the definition of bound and free variables:

\begin{Def}[Bound and Free variables]
  Given an OSDD $\psi=(s,k,Y)[\gamma_i :\psi_i]$ the bound variables
  of $\psi$, denoted $BV(\psi)$, are the output variables in $\psi$.
  The free variables of $\psi$, denoted $FV(\psi)$, are those
  variables which are not bound.
\end{Def}

Each OSDD corresponds to an MDD which can be constructed as follows. 

\begin{Def}[Grounding]\label{def:grounding}
  Given an OSDD $\psi=(s,k,Y)[\gamma_i:\psi_i]$, the MDD corresponding
  to it is denoted $\mathcal{G}(\psi)$ and is recursively defined as
  $\mathcal{G}(\psi)= (s,k,Y)
  [\alpha_j:\mathcal{G}(\psi_j[\alpha_j/Y])]$ where
  $\alpha_j \in type(Y)$ and $\psi_j = \psi_i$ such that
  $\gamma_i[\alpha_j/Y]$ is satisfiable.
\end{Def}

\begin{figure}[h]
  \centering
  \begin{tabular}{cp{0.0in}c}
    \begin{minipage}[t]{0.15\textwidth}
      \begin{adjustbox}{scale=0.75}
        $\vcenter{
        \begin{tikzpicture}
          \node[vertex] (a) {\texttt{msw(flip, 1, X1)}};
          \node[vertex] (b) [below=of a] {\texttt{msw(flip, 2, X2)}};
          \node[leaf] (s) [below left=of b] {$1$};
          \node[leaf] (f) [below right=of b] {$0$};
          \path[dir_edge] (a) -- (b);
          \path[dir_edge] (b) -- node[osddconstr] {$\mathtt{X1}=\mathtt{X2}$} (s);
          \path[dir_edge] (b) -- node[osddconstr] {$\mathtt{X1}\neq \mathtt{X2}$} (f);
        \end{tikzpicture}
        }$
      \end{adjustbox}
    \end{minipage}
    & &
        \begin{minipage}{0.8\textwidth}
          \centering
          \begin{adjustbox}{scale=0.6}
            $\vcenter{
            \begin{tikzpicture}
              \node[vertex] (a) {\texttt{msw(flip, 1, X1)}};
              \node[vertex] (bl) [below left=of a] {\texttt{msw(flip, 2, X2)}};
              \node[vertex] (br) [below right=of a] {\texttt{msw(flip, 2, X2)}};
              \node[leaf] (sl) [below left=of bl] {$1$};
              \node[leaf] (fl) [below right=of bl] {$0$};
              \node[leaf] (sr) [below left=of br] {$1$};
              \node[leaf] (fr) [below right=of br] {$0$};
              \path[dir_edge] (a) -- node[osddconstr] {$\mathtt{a}$} (bl);
              \path[dir_edge] (a) -- node[osddconstr] {$\mathtt{b}$} (br);
              \path[dir_edge] (bl) -- node[osddconstr] {$\mathtt{a}=\mathtt{X2}$} (sl);
              \path[dir_edge] (bl) -- node[osddconstr] {$\mathtt{a}\neq \mathtt{X2}$} (fl);
              \path[dir_edge] (br) -- node[osddconstr] {$\mathtt{b}=\mathtt{X2}$} (sr);
              \path[dir_edge] (br) -- node[osddconstr] {$\mathtt{b}\neq \mathtt{X2}$} (fr);
              
            \end{tikzpicture}
            }$
          \end{adjustbox}
        \end{minipage}\\
    (a) OSDD && (b) Grounding
  \end{tabular}
  \caption{Example of grounding OSDD}
  \label{fig:grounding}
\end{figure}
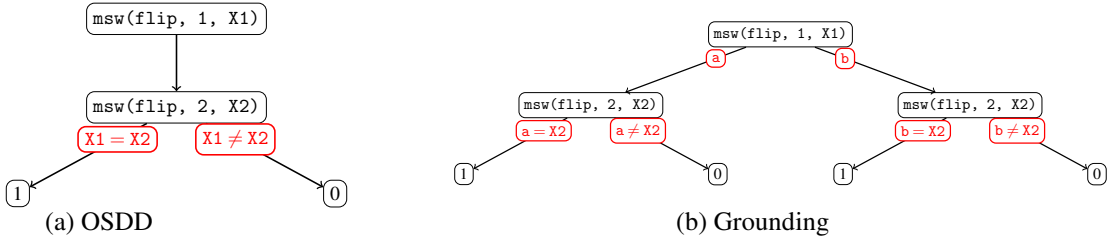

\begin{Ex}[Grounding]
  As an example consider a smaller version of the OSDD shown in
  Fig. \ref{fig:osdd-intro-ex}(a) as shown in
  Fig. \ref{fig:grounding}(a). In the first step of the grounding, all
  values satisfy the (empty) constraint of the outgoing
  edge. Therefore we get two subtrees which are identical except the
  substitution that is applied to the variable $X1$. In the next step
  (which we omit here), the subtrees get ground. Consider the left
  subtree, the value ``a'' satisfies the left branch, while the value
  ``b'' satisfies the right branch. Therefore in effect those edges
  get relabelled by ``a'' and ``b'' and similarly for the right
  subtree.
\end{Ex}

\myparagraph{Canonical OSDD representation.}  Given a total order
among variables and terms, the order of nodes in an OSDD is fixed.  We
can further order the outgoing edges uniquely based on the total order
of constraints labeling them.  This yields a canonical representation
of OSDDs.  In the rest of the paper we assume that OSDDs are in
canonical form.

\begin{Def}[Equivalence]
  \label{def:equivalence}
  All OSDD leaves which have the same node label are equivalent.  Two
  OSDDs $\psi = (s, k, Y)[\gamma_i : \psi_i]$ and $\psi' = (s, k,
  Y')[\gamma'_i : \psi'_i]$ are equivalent if $\forall i.\ \
  [\gamma_i:\psi_i] = [\gamma'_i:\psi'_i][Y/Y']$.
\end{Def}

We now define common operations over OSDDs which can create OSDDs from
primitives.

\begin{Def}[Conjunction/Disjunction]\label{def:conjunction-disjunction}
  Given OSDDs $\psi = (s, k, Y)[\gamma_i : \psi_i]$ and
  $\psi' = (s', k', Y')[\gamma'_j : \psi'_j]$, let $\oplus$ stand for
  either $\land$ or $\lor$ operation. Then $\psi \oplus \psi'$ is
  defined as follows.
  \begin{itemize}
  \item If $(s, k) \prec (s', k')$, then
    $\psi \oplus \psi' = (s, k, Y)[\gamma_i:\psi_i \oplus \psi']$
  \item If $(s', k') \prec (s, k)$ then
    $\psi \oplus \psi' = (s', k', Y')[\gamma'_j:\psi'_j
    \oplus \psi]$
  \item If $(s, k) = (s', k')$, first we apply the substitution
    $[Y/Y']$ to the second OSDD. Then
    $\psi\oplus\psi'=(s, k, Y)[\gamma_i\land\gamma'_j:\psi_i\oplus\psi'_j]$.
  \end{itemize}
\end{Def}

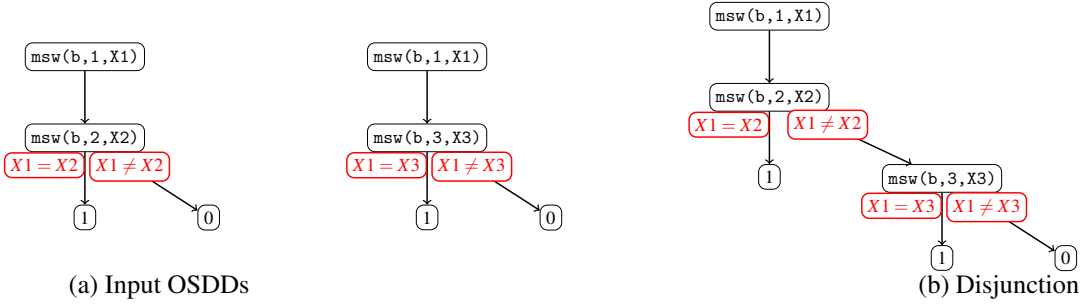
\begin{figure}
  \centering
  \begin{tabular}{c c c}
    \begin{minipage}[t]{0.25\textwidth}
      \begin{adjustbox}{scale=0.7}
        $\vcenter{
          \begin{tikzpicture}
            \node[vertex] (r1) {\texttt{msw(b,1,X1)}};
            \node[vertex] (c1) [below=of r1] {\texttt{msw(b,2,X2)}};
            \node[leaf] (s1) [below=of c1] {$1$};
            \node[leaf] (f1) [below right=of c1] {$0$};

            \path[dir_edge] (r1) -- (c1);
            \path[dir_edge] (c1) -- node[left, osddconstr] {$X1=X2$} (s1);
            \path[dir_edge] (c1) -- node[osddconstr] {$X1\neq X2$} (f1);
          \end{tikzpicture}
        }$
      \end{adjustbox}
    \end{minipage}
    &
      \begin{minipage}[t]{0.25\textwidth}
        \begin{adjustbox}{scale=0.7}
          $\vcenter{
            \begin{tikzpicture}
              \node[vertex] (r1) {\texttt{msw(b,1,X1)}};
              \node[vertex] (c1) [below=of r1] {\texttt{msw(b,3,X3)}};
              \node[leaf] (s1) [below=of c1] {$1$};
              \node[leaf] (f1) [below right=of c1] {$0$};
              
              \path[dir_edge] (r1) -- (c1);
              \path[dir_edge] (c1) -- node[left, osddconstr] {$X1=X3$} (s1);
              \path[dir_edge] (c1) -- node[osddconstr] {$X1\neq X3$} (f1);
            \end{tikzpicture}
          }$
        \end{adjustbox}
      \end{minipage}
    &
      \begin{minipage}[t]{0.5\textwidth}
        \begin{adjustbox}{scale=0.7}
          $\vcenter{
            \begin{tikzpicture}
              \node[vertex] (r1) {\texttt{msw(b,1,X1)}};
              \node[vertex] (c1) [below=of r1] {\texttt{msw(b,2,X2)}};
              \node[leaf] (s1) [below=of c1] {$1$};
              \node[vertex] (c2) [below right=of c1]
              {\texttt{msw(b,3,X3)}};
              \node[leaf] (s2) [below=of c2] {$1$};
              \node[leaf] (f1) [below right=of c2] {$0$};
              \path[dir_edge] (r1) -- (c1);
              \path[dir_edge] (c1) -- node[left, osddconstr] {$X1=X2$}
              (s1);
              \path[dir_edge] (c1) -- node[osddconstr] {$X1\neq X2$}
              (c2);
              \path[dir_edge] (c2) -- node[left, osddconstr] {$X1=X3$} (s2);
              \path[dir_edge] (c2) -- node[osddconstr] {$X1\neq X3$} (f1);
            \end{tikzpicture}
          }$
        \end{adjustbox}
      \end{minipage}\\
    (a) Input OSDDs & & (b) Disjunction
  \end{tabular}
  \caption{Disjunction of OSDDs}
  \label{fig:osdd-or}
\end{figure}

\begin{Ex}[Conjunction/Disjunction]
  Consider the input OSDDs in Fig. \ref{fig:osdd-or}(a). Their
  disjunction is shown in Fig. \ref{fig:osdd-or}(b). Disjunction of
  this OSDD with a third OSDD involving switch instance pairs $(b,2),
  (b,3)$ results in the OSDD shown in Fig. \ref{fig:osdd-intro-ex}(b).
\end{Ex}

Although OSDDs have been defined as trees, we can turn them into DAGs
by combining equivalent subtrees.  It is easy to generalize the above
operations to work directly over DAGs. The above operation may produce
improper OSDDs, but can be readily transformed to proper OSDDs as
follows 

\myparagraph{Transformation from improper to proper OSDDs.}
\begin{figure}[h]
	\centering
	\begin{tabular}{cp{0.0in}c}
		\begin{minipage}[t]{.35\textwidth}
			\begin{adjustbox}{scale=0.7}
				$\vcenter{
				\begin{tikzpicture}
				\node[vertex] (a) {\texttt{X}}; 
				\node[vertex] (b) [below=of a] {\texttt{Y}}; 
				\node[vertex] (c) [below=of b] {\texttt{Z}};
				\node[leaf] (d) [below left= 1.5cm and 3cm of c] {$\Psi_1$};
				\node[leaf] (e) [below left= 2cm and 1cm of c] {$\Psi_2$};
				\node[leaf] (f) [below left= 2.5cm and -1.75cm of c] {$\Psi_3$};
				\node[leaf] (fail1) [below left= 1cm and -2.5cm of c] {$0$};
				\path[dir_edge] (a) -- (b); 
				\path[dir_edge] (b) --  (c);
				\path[dir_edge] (c) -- node[osddconstr, below left= 0cm and 0cm] {$\mathtt{Z}=\mathtt{X} \land \mathtt{Z} = \mathtt{Y} $} (d);
				\path[dir_edge] (c) -- node[osddconstr, below left= .375cm and -.75cm] {$\mathtt{Z}=\mathtt{X} \land \mathtt{Z} \neq \mathtt{Y} $} (e);
				\path[dir_edge] (c) -- node[osddconstr, below right= .95cm and -.25cm] {$\mathtt{Z} \neq \mathtt{X} \land \mathtt{Z} = \mathtt{Y} $} (f);
				\path[dir_edge] (c) -- node[osddconstr, below right= 0cm and 0cm] {$\mathtt{Z} \neq \mathtt{X} \land \mathtt{Z} \neq \mathtt{Y} $} (fail1);
				\end{tikzpicture}
			}$
			\end{adjustbox}
		\end{minipage}
		& &
		\begin{minipage}{0.5\textwidth}
			\begin{adjustbox}{scale=0.75}
				$\vcenter{
				\begin{tikzpicture}
				\node[vertex] (a) {\texttt{X}}; 
				\node[vertex] (b) [below=of a] {\texttt{Y}}; 
				\node[vertex] (c) [below left= .75cm and 1cm of b] {\texttt{Z}};
				\node[vertex] (c2) [below right= .75cm and 2cm of b] {\texttt{Z}};
				
				\node[leaf] (d) [below left= 1.75cm and 0cm of c] {$\Psi_1$};
				\node[leaf] (fail1) [below right= 1cm and .5 cm of c] {$0$};
				
				\node[leaf] (e2) [below left= 1.75cm and 1cm of c2] {$\Psi_2$};
				\node[leaf] (f2) [below left= 2.5cm and -1cm of c2] {$\Psi_3$};
				\node[leaf] (fail2) [below left= 1cm and -2cm of c2] {$0$};
				
				\path[dir_edge] (a) -- (b); 
				\path[dir_edge] (b) -- node[osddconstr, below left= 0cm and 0cm] {$\mathtt{X}=\mathtt{Y} $} (c);
				\path[dir_edge] (b) --  node[osddconstr, below right= 0cm and 0cm] {$\mathtt{X} \neq \mathtt{Y} $} (c2);
				
				\path[dir_edge] (c) -- node[osddconstr, below left= 0.5cm and -.35cm] {$\mathtt{Z}=\mathtt{X} \land \mathtt{Z} = \mathtt{Y} $} (d);
				\path[dir_edge] (c) -- node[osddconstr, below right= 0cm and -.35cm] {$\mathtt{Z} \neq \mathtt{X} \land \mathtt{Z} \neq \mathtt{Y} $} (fail1);
				
				\path[dir_edge] (c2) -- node[osddconstr, below left= 0.5cm and -.25cm] {$\mathtt{Z}=\mathtt{X} \land \mathtt{Z} \neq \mathtt{Y} $} (e2);
				\path[dir_edge] (c2) -- node[osddconstr, below right = .95cm and -.8cm] {$\mathtt{Z} \neq \mathtt{X} \land \mathtt{Z} = \mathtt{Y} $} (f2);
				\path[dir_edge] (c2) -- node[osddconstr, below right= 0cm and 0cm] {$\mathtt{Z} \neq \mathtt{X} \land \mathtt{Z} \neq \mathtt{Y} $} (fail2);
				\end{tikzpicture}
			}$
			\end{adjustbox}
		\end{minipage}\\
		(a) Improper OSDD && (b) Proper OSDD
	\end{tabular}
	\caption{Transformation example}
	\label{fig:osdd-improper-to-proper}
\end{figure}
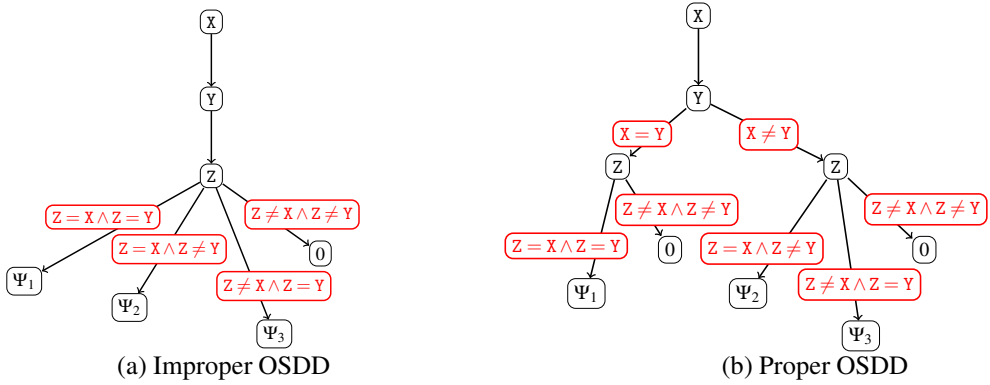

When performing and/or operations over proper OSDDs, the resulting
OSDD may be improper.  For instance, consider one OSDD with variables
$X,Z$, and another with variables $Y,Z$.  Constraints between
$X$ and $Z$, and those between $Y$ and
$Z$ may imply constraints between $X$ and
$y$ that may not be explicitly present in the resulting OSDD, thereby
violating the condition of explicit constraints
(Def. ~\ref{def:osdd}).

Fig.~\ref{fig:osdd-improper-to-proper}(a) shows an improper OSDD that
violates the explicit constraints condition.  For example, the edge
$(Z, \Psi_1)$, leading to the sub-OSDD $\Psi_1$, has edge constraints
which imply $X = Y$ while the edges leading to $\Psi_2$ and $\Psi_3$
imply $X \neq Y$.

An improper OSDD can be converted into a proper one by rewriting it
using a sequence of steps as follows.  First, we identify an implicit
constraint and where it may be explicitly added without violating the
urgency property.  In the example, we identify $X=Y$ as implicit, and
attempt to introduce it on the outgoing edges of node labeled $Y$.
This introduction splits the edge from $Y$ to $Z$ into two: one
labeled $X=Y$, and another labeled $X\neq Y$, the negation of the
identified constraint.  The original child rooted at $Z$ is now
replicated due to this split.  We process each child, eliminating
edges and corresponding sub-OSDDs whenever the constraints are
unsatisfiable.  In the example in
Fig.~\ref{fig:osdd-improper-to-proper}(b) we see that the edges
$(Z, \Psi_2)$ and $(Z, \Psi_3)$ have been removed from $\Psi_{Z_1}$
since their constraint formula are inconsistent with $X = Y$.  We
repeat this procedure until no implicit constraints exist in $\Psi$.

\comment{ Similar to BDDs we can define \emph{reduced} OSDDs by
  eliminating nodes whose children are all equivalent; however, a node
  may be eliminated only if its output variable is not used in a
  constraint in some descendant subtree, as defined below:

\begin{Def}[Marginalization]\label{def:marginalization}
  Given an OSDD $\psi = (s, k, Y)[\gamma_i : \psi_i]$ such that
  $\forall i. \psi_i = \psi'$ and $Y \not \in FV(\psi')$, then $\psi =
  \psi'$
\end{Def}
}


\myparagraph{Constraint application.} An OSDD may be specialized for a
context which constrains its variables, as follows.
\begin{Def}[Constraint application]\label{def:constraint-apply}
  Given an OSDD $\psi = (s,k,Y)[\gamma_i:\psi_i]$ and an atomic
  constraint $\beta$, the application of $\beta$ to $\psi$ results in a
  new OSDD $\psi'$ as follows:
  \begin{itemize}
  \item Application of $\beta$ to $0, 1$ yields $0, 1$ respectively.
  \item If $Vars(\beta) \subseteq \mathcal{O}(n)$ where $n$ is the root of
    $\psi$, then $\psi' = (s,k,Y)[\gamma_i \land \beta: \psi_i, \neg
    \beta: 0 ]$
  \item Else $\psi' = (s,k,Y)[\gamma_i : \psi'_i]$ where each
    $\psi'_i$ results from the application of $\beta$ to $\psi_i$.
  \end{itemize}
\end{Def}

\myparagraph{Properties.}  OSDDs and the operations defined above have
a number of properties necessary for their use in representing
explanations for query evaluation in PLPs.

\begin{Pro}[Closure]\label{pro:closure}
  OSDDs are closed under conjunction and disjunction
  operations.
\end{Pro}

The following shows that conjunction and disjunction operations over
OSDDs lift the meaning of these operations over ground MDDs.
\begin{Pro}[Compatibility with Grounding]
  Let $\psi = (s, k, Y)[\gamma_i : \psi_i]$ and
  $\psi' = (s', k', Y')[\gamma'_j : \psi'_j]$ be two OSDDs, then
  \begin{gather*}
    \mathcal{G}(\psi \oplus \psi') = \mathcal{G}(\psi) \oplus
    \mathcal{G}(\psi').
  \end{gather*}
\end{Pro}


\section{Construction}
\label{sec:construction}

Given a definite PLP program and a ground query,  we construct an OSDD
as the first step of the inference process.  The construction is done
via constraint-based tabled evaluation over a transformed program.  At
a high level, each $n$-ary predicate $p/n$ in the original PLP
program is transformed into a $n+2$-ary predicate $p/(n+2)$ with one
of the new arguments representing an OSDD at the time of call to $p$, and
the other representing OSDDs for answers to the call.

For simplicity, although the transformed program represents OSDDs as Prolog terms, we
reuse the notation from Section~\ref{sec:osdd} to describe the
transformation.

\myparagraph{Transformation.}
We use $\overline{T_1}, \overline{T_2}, 
\ldots$ to represent tuples of arguments.
Clauses in a definite program may be of one of two forms:
\begin{itemize}
\item Fact $p(\overline{T})$: is transformed to another fact of the
  form $p(\overline{T}, \mathtt{O}, \mathtt{O})$, denoting that a fact
  may bind its arguments but do not modify a given OSDD.
\item Clause $\id{head} \leftarrow \id{body}$:  without loss of
  generality, we assume that the body is binary: i.e., clauses are of
  the form $p(\overline{T}) \leftarrow q(\overline{T_1}),
  r(\overline{T_2})$.  Such clauses are transformed into
  $p(\overline{T}, \mathtt{O_1}, \mathtt{O_3}) \leftarrow
  q(\overline{T_1}, \mathtt{O_1}, \mathtt{O_2}),
  r(\overline{T_2}, \mathtt{O_2}, \mathtt{O_3})$
\end{itemize}

For each user-defined predicate $p/n$ in the input program, we add the
following directive for the transformed predicate $p/(n+2)$
\begin{quote}
  \texttt{:- table $p(\_, \ldots, \_,$ lattice(or/3)).}
\end{quote}
which invokes answer subsumption~\cite{Swift2010} to group all answers
by their first $n+1$ arguments, and combine the $n+2$-nd argument in
each group using \texttt{or/3}, which implements the disjunction
operation over OSDDs.

\myparagraph{Constraint-Based Evaluation.}
An important aspect of OSDD construction is constraint processing.
Our transformation assumes that constraints are associated with
variables using their \emph{attributes}~\cite{attr}.  We assume the
existence of the following two
predicates:
\begin{itemize}
\item \texttt{inspect(X, C)}, which, given a variable \texttt{X},
  returns the constraint formula associated with \texttt{X}; and
\item \texttt{store(C)}, which, given a constraint formula \texttt{C},
  annotates all variables in \texttt{C} with their respective atomic
  constraints.
\end{itemize}
For tabled evaluation, we assume that each table has a local
constraint store~\cite{SR:PADL07}. Such a constraint store can be
implemented using the above two predicates.

\myparagraph{OSDD Builtins.}
The construction of the transformed program is completed by defining
predicates to handle the two constructs that set PLPs apart:
\begin{itemize}
\item \texttt{msw(S, K, X, $\mathtt{O_1}$,  $\mathtt{O_1}$)}: Note
  that \texttt{msw}'s in the body of a clause would have been
  transformed to a 5-ary predicate.  This predicate is defined as:
  \begin{align*}
    msw(S,K,Y,\mathtt{O_1}, \mathtt{O_2}) \leftarrow & 
    \mathtt{inspect(}Y, \gamma\mathtt{)},\\
                            & (\gamma = \{\} \rightarrow O = (S,K,Y)[\{\} : 1]\\
                            & ; \ \  O = (S,K,Y)[\gamma: 1, \neg \gamma: 0]\\
                            &), \mathtt{and(O_1, O, O_2)}.
  \end{align*}
  where \texttt{and/3} implements conjunction operation over OSDDs.
\item Constraint handling: constraints in the input program will be
  processed using: 
  \begin{align*}
    constraint(C, \mathtt{O_1}, \mathtt{O_2}) \leftarrow 
    & ((\id{Vars}(C) \cap BV(\mathtt{O_1})) \not = \emptyset\\
    & \rightarrow  \mathtt{O_2} = \mathtt{applyConstraint}(C, \mathtt{O_1})\\
    & ; \ \    \mathtt{O_2} = \mathtt{O_1}\\
    &  ), \mathtt{store}(C).
  \end{align*}
  where \texttt{applyConstraint} is an implementation of Defn.~\ref{def:constraint-apply}
\end{itemize}

To compute the OSDD for a ground atom $q(\overline{X})$ in the
original program, we evaluate $q(\overline{X},1,O)$ to obtain the
required OSDD as $O$.

\section{Inference}
\label{sec:inference}

\myparagraph{Exact Inference.}  Given an OSDD
$\psi=(s,k,Y)[\gamma_i:\psi_i]$, let $Dom(FV(\psi))$ be the Cartesian
product of the types of each $X \in FV(\psi)$. We define a function
$\pi$ which maps an OSDD $\psi$ and a substitution
$\sigma \in Dom(FV(\psi))$ to $\mathbb{R}$. Its definition for leaves
is as follows
\begin{gather*}
  \pi(1,\emptyset) = 1 \text{ and }
  \pi(0,\emptyset) = 0.
\end{gather*}
Next for an OSDD $\psi$ and an arbitrary substitution $\sigma'$,
define $\Pi(\psi,\sigma')$ to be
\[
  \Pi(\psi,\sigma') = \sum_{\sigma : \sigma \nparallel \sigma'} \pi(\psi, \sigma),
\]
where constraint formulas $\sigma$ and $\sigma'$ are said to be
compatible if their conjunction is satisfiable (denoted
$\sigma \nparallel \sigma'$).  For internal nodes we define
$\pi(\psi,\sigma)$ as
\[
  \pi(\psi,\sigma) = \sum_i \sum_{y \in \sem{\gamma_i\sigma}_Y} P(Y = y) \Pi(\psi_i,\sigma[y/Y]).
\]
where $P(Y=y)$ is the probability that $k$-th instance of $s$ has
outcome $y$.

Given a ground query whose OSDD is $\psi$ we return the answer
probability as $\Pi(\psi,\emptyset)$.

\begin{Pro}[Complexity]
  \label{pro:complexity}
  The time complexity of probability computation of OSDD $\psi$ is
  $O(D\cdot N \cdot exp(V))$ where $D$ is the maximum cardinality of
  all types, $N$ is the number of nodes in $\psi$, and $V$ is the size
  of the largest set of free variables among all internal nodes of
  $\psi$.
\end{Pro}

Under certain conditions we can avoid the exponential complexity of
the naive probabilistic inference algorithm.  By exploiting the
regular structure of the solution space to a constraint formula we
avoid the explicit summation $\sum_{y \in \sem{\gamma_i\sigma}_Y}$.
In this case we say that $\gamma_i$ is measurable.  We formally define
measurability and a necessary and sufficient condition for
measurability.

\begin{Def}[Measurability]
  \label{def:measurability}
  A satisfiable constraint formula $\gamma$ is said to be measurable
  w.r.t $X \in Vars(\gamma)$ if for all ground substitutions $\sigma$
  on $Vars(\gamma)\setminus \{X\}$ which satisfy $\gamma$,
  $|\sem{\gamma\sigma}_X|$ is equal to a unique value $m_X$ called the
  measure of $X$ in $\gamma$.
\end{Def}

\begin{Def}[Saturation]
  \label{def:property1}
  A constraint formula $\gamma$ is said to be saturated if its
  constraint graph satisfies the following condition: For every
  $X \in Vars(\gamma)$, let $\mathcal{Z}$ be the set of nodes
  connected to $X$ with a $``\neq"$ edge. Then there exists an edge
  ($``\neq"$ or $``="$) between each pair of nodes in $\mathcal{Z}$
  (except when both nodes in the pair represent constants).
\end{Def}

\begin{Pro}[Condition for Measurability]
  A satisfiable constraint formula is measurable w.r.t all of its
  variables if and only if it saturated.
\end{Pro}

\begin{Def}[Measurability of OSDDs]
  \label{def:measurability-osdds}
  An OSDD is said to be measurable, if for each internal node $n$ and
  outgoing edge labeled $\gamma_i$, the constraint formula obtained by
  taking the conjunction of $\gamma_i$ with the constraint formula on
  the path from root to $n$ is measurable w.r.t the output variable in
  node $n$.
\end{Def}

\begin{Ex}[Measurability]
As an example, consider the OSDD for the birthday problem. It
satisfies the measurability condition. The measures for the constraint
formulas labeling the edges are shown in
Fig. \ref{fig:osdd-intro-ex}(b).
\end{Ex}

When an OSDD is measurable and all distributions are uniform, the
probability computation gets specialized as follows:
\[
  \pi(\psi,\sigma) = \sum_{i} m_i P(Y=\hat{y_i}) \Pi(\psi_i,\sigma[\hat{y_i}/Y])
\]
where $\hat{y_i}$ is an arbitrary value in $\sem{\gamma_i\sigma}_Y$
and $P(Y=\hat{y_i}) = \frac{1}{|type(Y)|}$.

\begin{Pro}[Complexity for Measurable OSDDs]
  If an OSDD $\psi$ is measurable and all switches have uniform
  distribution, the complexity of computing probability of $\psi$ is
  $O(D\cdot N)$ where $D$ is the maximum cardinality of all types and $N$
  is the number of nodes in $\psi$.
\end{Pro}

\myparagraph{Likelihood Weighted Sampling.}
Likelihood weighting is a popular sampling based inference technique
in BNs. The technique can be described as follows: Sample all
variables except evidence variables in the topological order. The
values of evidence variables are fixed and each one of them
contributes a \emph{likelihood weight}, which is the probability of its
value given the values of its parents. The likelihood weight of the
entire sample is the product of the likelihood weights of all evidence
variables. This technique has been shown to produce sample estimates
with lower variance than independent sampling
\cite{fung1990weighting,shachter1990simulation}. 

Likelihood weighted sampling can be generalized to PLPs as follows:
Given an OSDD $\psi=(s,k,Y)[\gamma_i:\psi_i]$ for evidence, generate a
sample as follows:
\begin{itemize}
\item Construct $type'(Y) = type(Y) \setminus \cup_j\sem{\gamma_j}_Y$ where
  $\psi_j = 0$.
\item If $type'(Y) = type(Y)$ sample $y$ from the distribution of $Y$
  leaving likelihood weight of the sample unchanged.
\item Otherwise, sample $y$ uniformly from $type'(Y)$ and multiply the
  likelihood weight of the sample by $P(Y=y)$. Let
  $ y \in \sem{\gamma_i}_Y$ for some $i$. Then continue construction
  of the sample by recursively sampling from the OSDD $\psi_i[y/Y]$
\end{itemize}

For PLPs encoding BNs and MNs, the simple nature of evidence allows us
to generate only consistent samples. However, for general queries in
PLP, it is possible to reach a node whose edge constraints to non-$0$
children are unsatisfiable. In that case, we reject the current sample
and restart. Thus, we generalize traditional LW sampling.

To compute conditional probabilities, samples generated for evidence
are extended by evaluating queries. The conditional probability of the
query given evidence is computed as the sum of the likelihood weights
of the samples satisfying the query and evidence divided by the sum of
the likelihood weights of the samples which satisfy evidence. To
compute unconditional probability of a query, we simply compute the
average likelihood weight of the samples satisfying the query.

\begin{Ex}[Likelihood weighting]
  Consider the OSDD shown in Fig. \ref{fig:osdd-intro-ex}(a). To
  generate a likelihood weighted sample we start from the root and
  sample the random variables. The first three nodes do not have any
  constraints on their outgoing edges, therefore we can sample those
  random variables from their distributions. Assume that we get the
  sequence ``aba''. The likelihood weight of the sample remains 1 at
  this stage. When sampling the random variables at the next three
  nodes, $type'(Y)$ gets restricted to a single value. Since the
  distributions are all uniform the likelihood of the entire sample
  becomes $0.5^3$. All samples would have the same likelihood weight
  and therefore the probability of ``evidence(6)'' is $0.125$
\end{Ex}

\comment{
$\psi$ for the evidence goal. We then generate likelihood weighted (LW)
samples by traversing $\psi$ top-down, using an initial likelihood
weight of $1$. A node $(s(\overline{X}),k,Y)$ with no $0$ child means
that there is a possibility of success for each valuation of $Y$.  We
then sample $Y$ from $s(\overline{X})$'s declared distribution, leaving the
likelihood weight unchanged, and choose the child subtree based on the value of $Y$ value to
continue the sample.  A node $(s(\overline{X}),k,Y)$ with a $0$ child
means that failure is possible for certain valuations of $Y$.  In this
case, we first find the set $S$ of all values of $Y$ that are
consistent with an edge constraint that leads to a non-$0$ node.  We
sample $Y$ uniformly from $S$, and multiply the likelihood weight with
the probability of $Y$ in $s(\overline{X})$'s declared distribution.  Sampling using a uniform
distribution is not the only choice for sampling from the set of
allowed values. We can use special distributions which reduce the
variance of likelihood weights (i.e, importance sampling). An example
is given in Section~\ref{sec:expt}.  A sample for some conditioning evidence is
completed when a $1$ node is reached.

Note that since the
constraint propagation methods used to build the OSDD are inexact
(short of satisfiability), it is possible to reach a node whose edge
constraints to non-$0$ children are unsatisfiable.  In that case, we
reject the current sample and restart.  Thus, we generalize
traditional LW sampling when limitations of constraint propagation
proscribe guaranteed generation of samples consistent with evidence.

Each LW sample is extended by evaluating the query goal.  During the
second step, when we see an \texttt{msw} for a process/instance not
encountered before, its outcome is independently sampled from the
process's declared distribution. The sum of the likelihood weights of
samples which satisfy the query and evidence divided by the sum of the
likelihood weights of the samples which satisfy evidence gives an
estimate of the conditional probability. For unconditional
probability, we build OSDD $\psi$ for the query goal, and estimate the
probability as the average likelihood weight of the samples.
}


\section{Experimental Evaluation}
\label{sec:expt}

We present the results of experiments using a prototype implementation
of a likelihood weighted sampler based on OSDDs. The prototype uses
XSB Prolog to construct OSDDs, and a few modules written in C for
maintaining the sampler's state and dealing with random variable
distributions.  We used the following examples in the experiments.

\begin{figure}
\centering
\begin{tabular}{cc}
  \comment{
\mbox{
\includegraphics[width=2.7in]{gridanswer.eps}}
& 
\mbox{
\includegraphics[width=2.7in]{isinganswer.eps}}\\
  (a) BN & (b) Ising Model\\
  }
\mbox{
\includegraphics[width=2.7in]{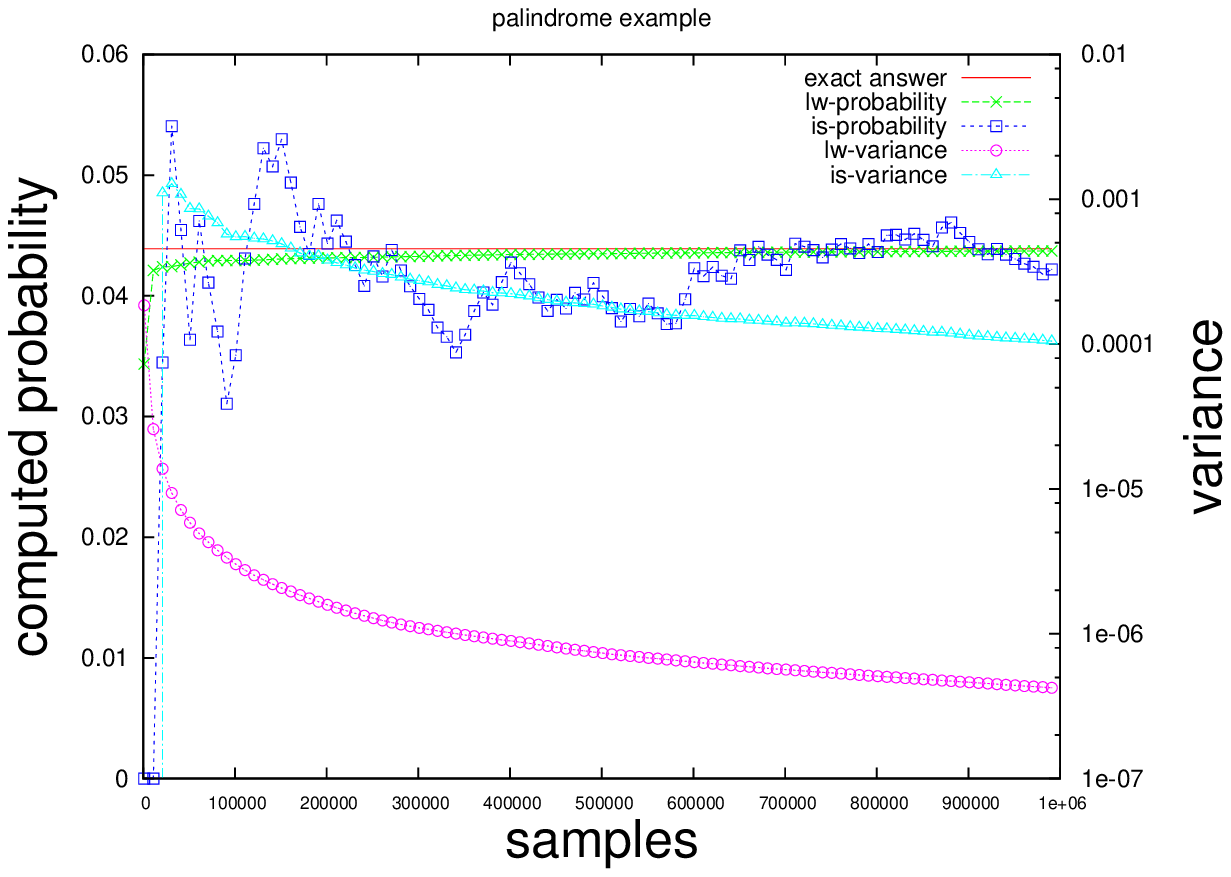}}
& 
\comment{
\mbox{
\includegraphics[width=2.7in]{gradesanswer.eps}}\\}
\mbox{
\includegraphics[width=2.7in]{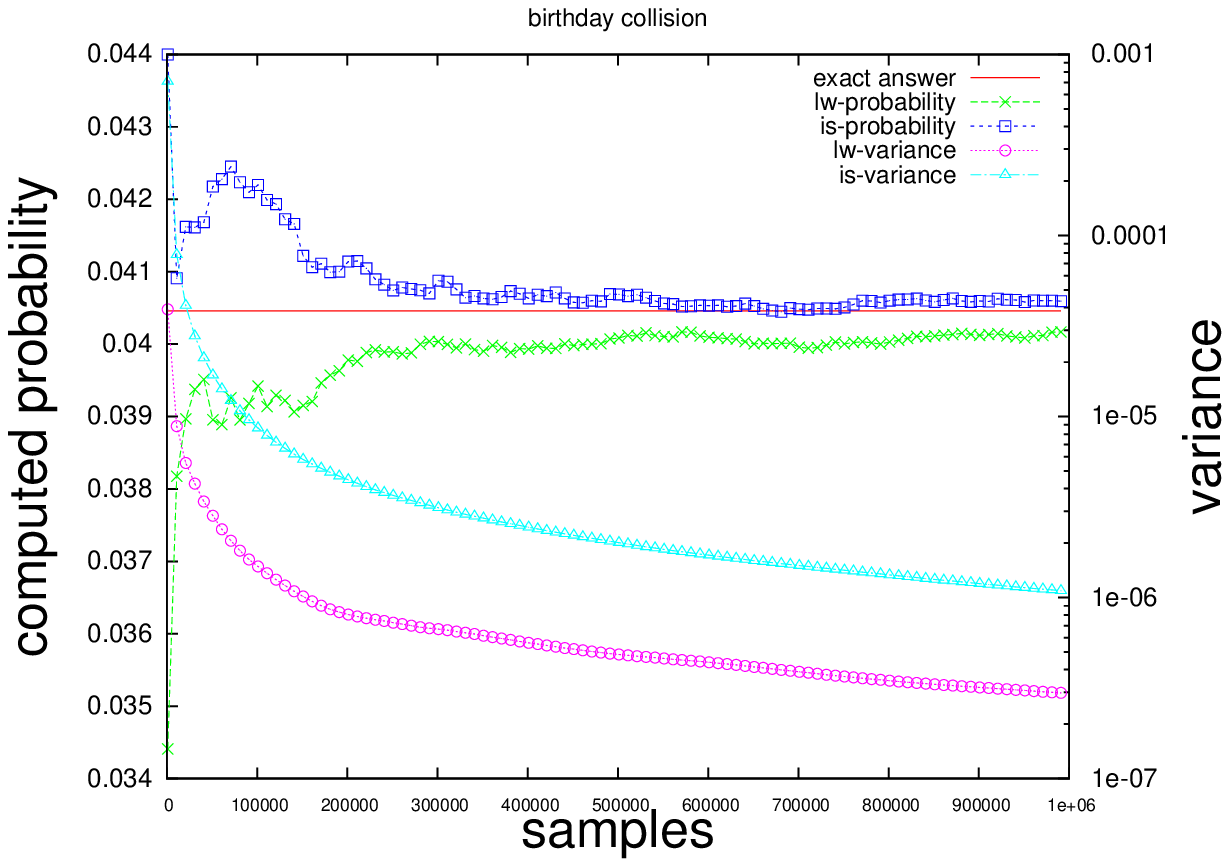}}\\
\comment{(c) Palindrome & (d) PRM Model}
(a) Palindrome & (b) Birthday
\end{tabular}
\caption{Experimental Results}
\label{fig:results1}
\vspace*{-.2in}
\end{figure}

\begin{itemize}
  \comment{
\item \textbf{Grid BN} is a Bayesian Network with Boolean random
  variables arranged in a $6\times6$ grid (with arcs going
  left-to-right and top-to-bottom). This structure was used to
  evaluate the effectiveness of our technique when the evidence
  probability is extremely low (\string~$10^{-12}$).
\item \textbf{Ising Model} is a well-known undirected graphical
  model.  We used a $6\times6$ grid of Boolean random variables with
  factors on edges. The PRISM program independently generates values
  of terminal nodes of all edges, and ties them together by
  expressing equality constraints between shared variables of
  edges.  }
\item \textbf{Palindrome}, which is shown in
  Fig.~\ref{fig:intro-ex-palindrome}, with evidence limited to strings
  of length $N=20$, and query checking $K=4$
  ``\texttt{a}''s. While the compuation of evidence probability is
  easy, the computation of the probability for conjunction of query
  and evidence is not. This is because, the query searches for all
  possible combinations of 'K' positions. Therefore, for large problem
  sizes, likelihood weighting has to be used to get approximate
  answers.

\comment{\item \textbf{Course Grades PRM} is a standard 
PRM example \cite{heckerman2004probabilistic} which
models the grades obtained by students in various courses.  
Due to the relational structure in
the PRMs, queries can have multiple symbolic derivations.}

\item \textbf{Birthday}, shown in Fig.~\ref{fig:intro-ex-birthday}
   with population size of $6$, i.e.  query
  \texttt{same\_birthday(6)}. While this query can be evaluated by 
  exact inference due to measurability, we use it to
  test the performance of likelihood weighted sampler.
\end{itemize}
\comment{The PRM example also had a conditional query
but with evidence of reasonable likelihood.}
\comment{The first three examples had a
single symbolic derivation while the last two examples had more
than one symbolic derivation.}  \comment{It should be noted that only the first
example, \textbf{Grid BN}, can be evaluated in the PRISM system; the
other examples have queries that violate PRISM's mutual exclusion and
independence assumptions and hence cannot be directly evaluated in
that system.   Our inference procedure, however, removes PRISM's
assumptions and correctly evaluates the query probabilities for all
the above examples.}
The results of the experiments are shown in
Fig.~\ref{fig:results1}.
\comment{Figs.~\ref{fig:results1} and ~\ref{fig:results2}.}  Each subfigure plots the estimated
probability and variance of the estimate (on log scale), for two samplers: the LW
method described in this paper, and a simple independent sampler (with
rejection sampling for conditional queries).  Note that the LW
sampler's results show significantly lower variance in both the examples.
\comment{For the Grid BN and Ising Model, the evidence
probability was low enough that a rejection sampler was unable to
draw a consistent sample.  The LW sampler, however, was able to
converge to a reasonable estimate of low variance in about 500,000
samples.}
\comment{
Both examples generated a single symbolic derivation.  We
directly sampled from this instead of materializing an OSDD structure.
For the Grid BN, node consistency was sufficient to derive
domain restrictions.}
\comment{For the Ising model, we found that standard LW sampling
(picking a restricted value uniformly and assigning a likelihood
weight) generated a number of samples with extremely low
weights. Instead the probabilites of the set of allowed values were
normalized to create a new proposal distribution. This resulted in
generating samples with higher likelihood weights.}

\comment{For both Palindrome and PRM examples, the LW sampler quickly converges
  to the actual probability.}

For the Palindrome example, the LW sampler quickly converges to the
actual probability, while the independent sampler fails to converge
even after a million samples.  The unusual pattern of variance for
independent sampler in the initial iterations is due to it not being
able to generate consistent samples and hence not having an estimate
for the answer probability. \comment{For the PRM example, node
  consistency was sufficient to propagate constraints within a single
  derivation. } \comment{
\begin{wrapfigure}{r}{0.48\textwidth}
\vspace*{-.1in}
\mbox{
\includegraphics[width=2.7in]{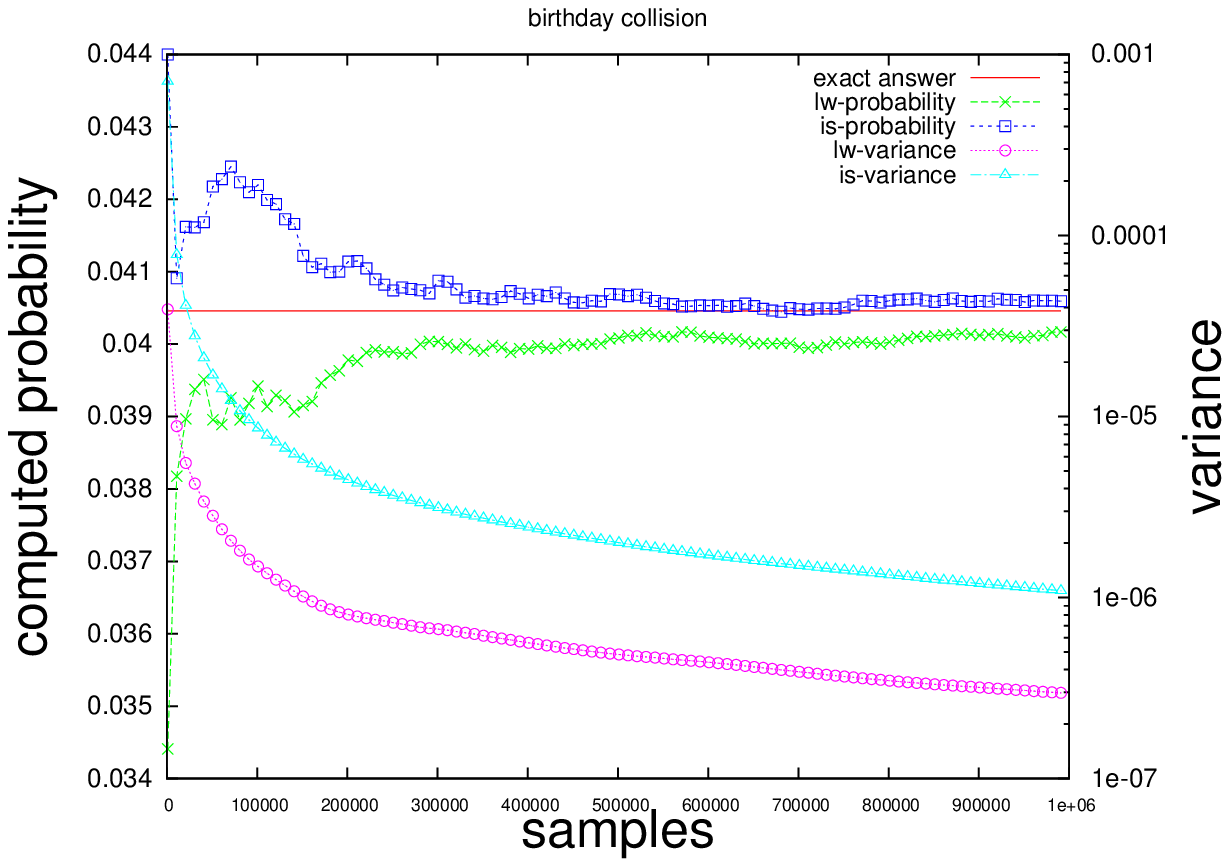}}
\caption{Results for Birthday Collision Example}
\label{fig:results2}
\vspace*{-.1in}
\end{wrapfigure}
}
In the birthday example, we notice that all consistent samples have
the same likelihood weight and they are quite low. Due to this reason,
likelihood weighting doesn't perform much better than independent
sampling. Interestingly, 
independent sampling using the OSDD structure was significantly faster
(up to $2\times$) than using the program directly.  This is because
the program's non-deterministic evaluation has been replaced by a
deterministic traversal through the OSDD.

\comment{
\subsection{Grid BN}
This experiment mainly demonstrates the effectiveness of constraint
propagation in the context of extremely low probability evidence. The
PGM is a 6x6 grid shaped BN. All the random variables are boolean. The
CPTs and the evidence are chosen to give extremely low probability to
evdience (1e-12). Due to the extremely skewed nature of the CPTs
independent sampler fails to find consistent samples even after a
million attempts.  In contrast, likelihood-weighted sampler can
generate all consistent samples using node consistency algorithm
alone. When the probabilities are not so skewed, then both independent
sampler and likelihood-weighted sampler are expected to converge in
comparable number of samples. The results on grid BN are shown in
Fig~\ref{fig:gridlw}.
}

\comment{
\subsection{Course grades PRM}
This is a standard PRM example \cite{heckerman2004probabilistic} which
models the grades obtained by students in various courses. PRMs can
define random variables over a population and specify independencies
between them without constructing a ground network. As such they are
more expressive than ordinary BNs. Due to the relational structure in
the PRMs, queries can have multiple symbolic derivations (in contrast
to queries on BNs). Node consistency is sufficient to propagate
constraints within each symbolic derivation. However, to handle the
disjunctive constraints we use the idea of constructive disjunction
\cite{van1998design}. The conditional query we tested was the grade of
a particular student-course pair, given the grades of other
student-course pairs. The results are shown in Fig~\ref{fig:grades}.
\begin{figure}
\centering
\includegraphics[width=3in]{gradesanswer.eps}
\caption{Course grades PRM}
\label{fig:grades}
\end{figure}
}

\comment{
\subsection{Ising Models}
Ising models are well known examples of PGMs where exact inference is
intractable. Ising models along with the general class of Markov
networks can be encoded in PLPs by sampling edges using switches and
expressing equality constraints between shared variables of
edges. Independent sampling fails to generate consistent samples for
even moderate ($6\times6$) sized Ising models. In the case of
likelihood-weighted sampler, node consistency and arc consistency do
not yield any refinement of the domains. Forward-checking on the other
hand will generate consistent samples without any failures. Yet the
samples generated by the naive likelihood weighting approach have
extremely low weights and do not contribute to meaningful
estimates. The problem is that for an edge which has one of its
variables restricted, there are two possible legal values: one which
gives same spins to both nodes and the other giving opposite
spins. Likelihood weighted sampler samples these uniformly, thereby
generating very low weight samples. The solution is to sample from the
legal values using a renormalized distribution. The revised approach
gives much improved estimates. The results for 6x6 ising model where
the query was the probability of one of the stable configuration are
shown in Fig~\ref{fig:ising}.
\begin{figure}
\centering
\includegraphics[width=3in]{isinganswer.eps}
\caption{Ising model}
\label{fig:ising}
\end{figure}
}

\comment{
\subsection{Birthday Collision}
In this experiment we consider the birthday collision problem: given a
set of randomly chosen people, what is the probability that atleast
two of them have the same birthday. Even though this problem can be
solved analytically, generalizations where birthday is chosen from
some distribution other than the uniform distribution cannot be solved
analytically. We considered the problem with six persons. Neither node
consistency nor arc-consistency given any reduction in the
domains. Moreover there are multiple symbolic derivations. Therefore
likelihood-weighted sampler uses forward checking together with
constructive disjunction. The results are shown in
Fig~\ref{fig:birthday}. We observe that likelihood-weighted sampler
has no significant advantage over independent sampler. This can be
understood by the fact that only the last variable to be sampled
(i.e., birthday of last person) has its domain restricted. Even though
inconsistent samples are avoided, the generated samples have low weight.
\begin{figure}
\centering
\includegraphics[width=3in]{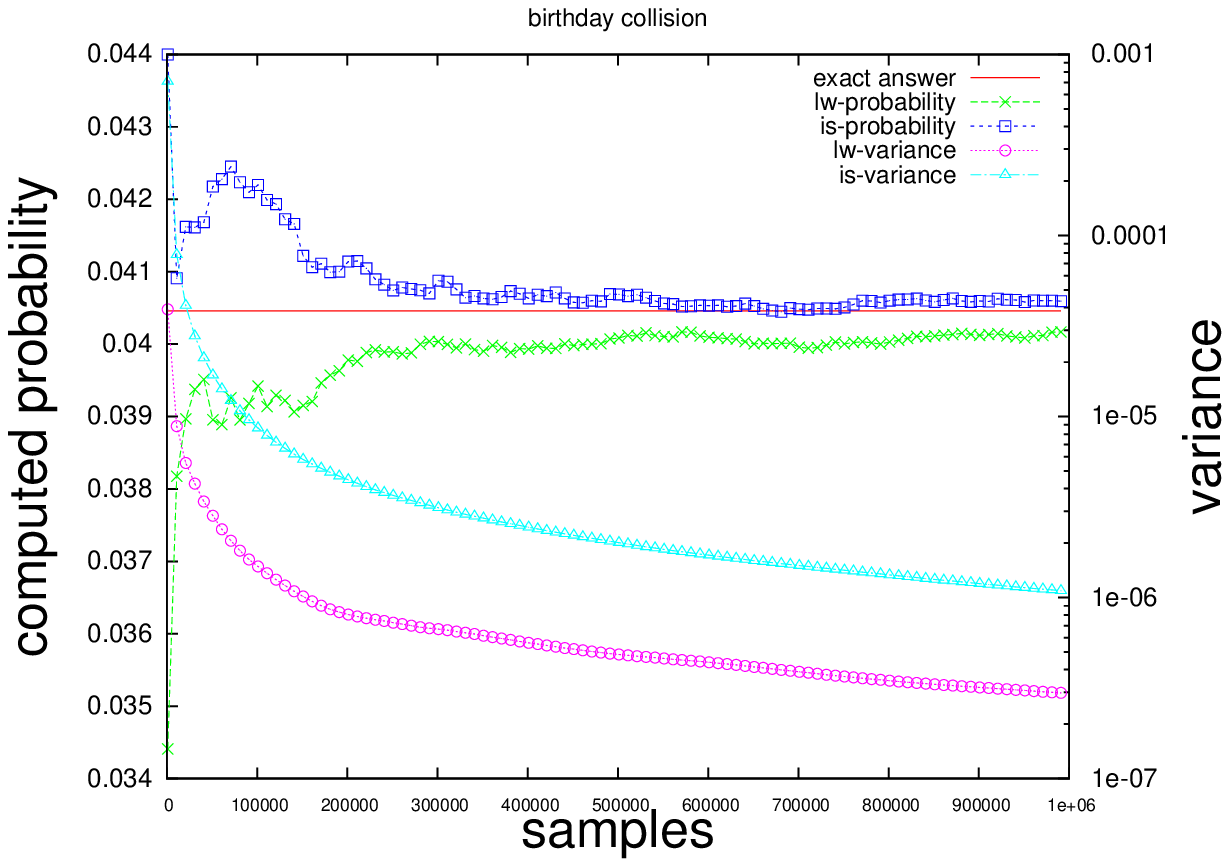}
\caption{Birthday collision}
\label{fig:birthday}
\end{figure}
}

\comment{
\subsection{Palindrome}
In this experiment we considered a two state Markov chain (states A and
B) with uniform transition probabilities. The evidence is that a fixed
length state transition sequence of this Markov chain is a
palindrome. The query is the probability of a certain number of
occurrences of state A. This probability can be computed analytically
in the case of two states and uniform distributions. However, when the
transition probabilities are made non-uniform, it is not amenable to
analytical solution. The convergence of likelihood weighted sampler
and independent sampler for the query of four As in a palindrome of
length ten is shown in Fig~\ref{fig:palindrome}.
\begin{figure}
\centering
\includegraphics[width=3in]{palindrome4-10.eps}
\caption{Palindrome}
\label{fig:palindrome}
\end{figure}
}

\comment{
\paragraph{Overheads.}
For all the examples, the time to construct the OSDDs, was negligible
(ranging from 4ms for Grid BN to 7ms for Birthday example, with XSB
3.5.0 on a 2.5GHz Intel Core 2 Duo machine).  However, while an independent sampler
picks values from the given distributions, the likelihood-weighting
sampler needs to construct restricted domains to draw samples from.
Consequently, the LW sampler takes up to $4\times$ per sample as an
independent sampler.
}

\begin{figure}[h]
  \centering
  \begin{tabular}{c c}
    \comment{
    \mbox{
    \includegraphics[width=0.3\linewidth]{gridbn-problog}}
    &}
    \mbox{
    \includegraphics[width=0.5\linewidth]{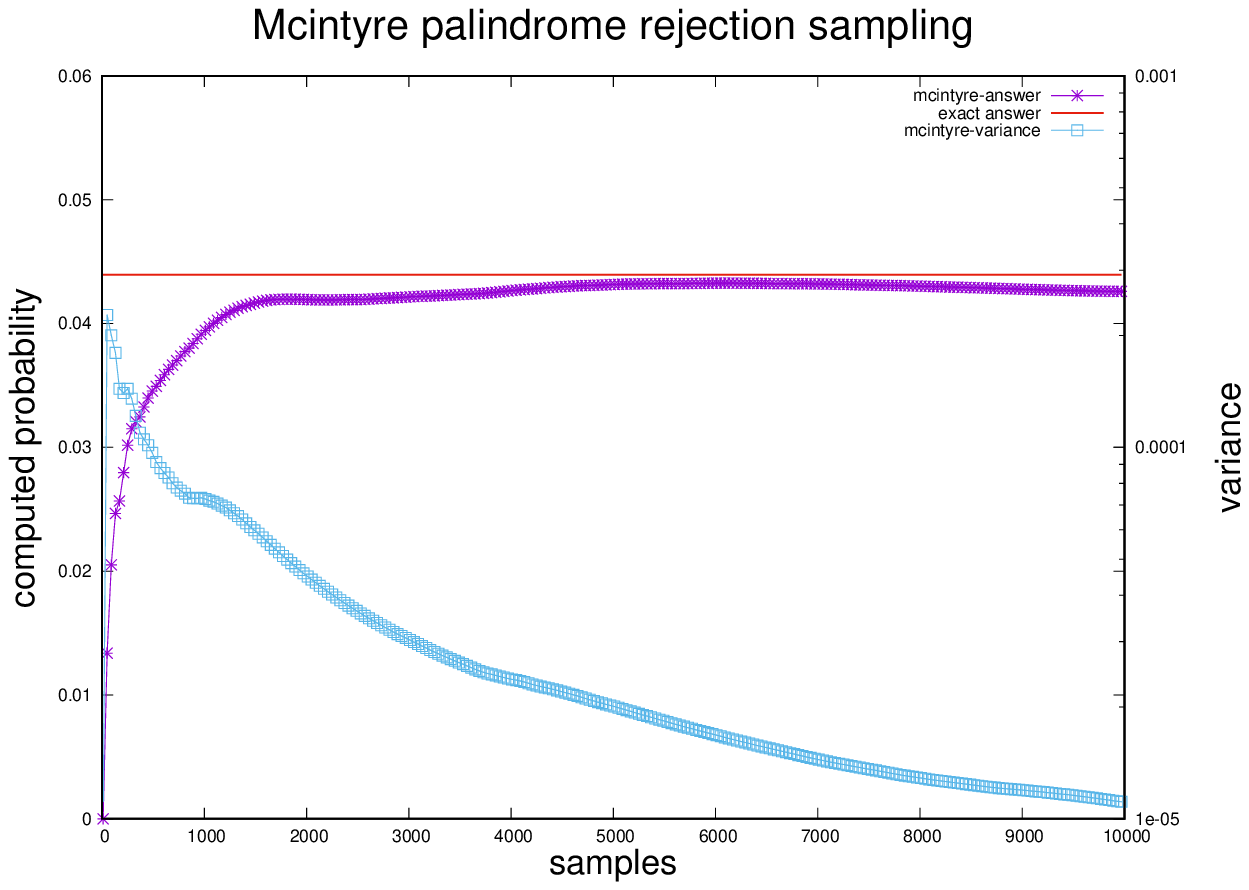}}
    &
    \mbox{
    \includegraphics[width=0.5\linewidth]{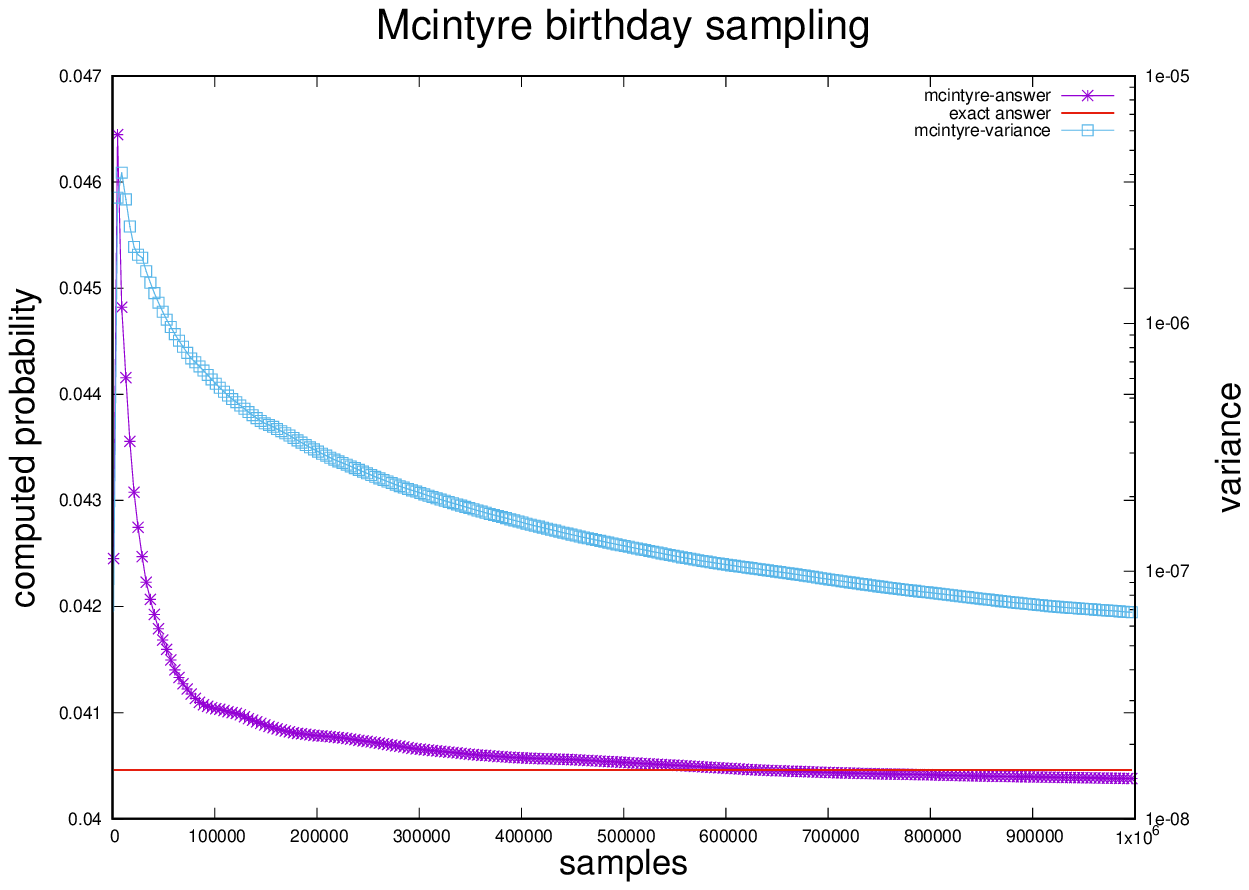}}\\
    PITA palindrome & PITA birthday\\
  \end{tabular}
  \caption{Experimental Results for PITA samplers}
  \label{fig:pita-problog-plots}
\end{figure}

\paragraph{Comparison with PITA and ProbLog samplers}
\label{para:other-samplers}
We compared the performance of LW sampler with the sampling based
inference of ProbLog and PITA \footnote{We used a core i5 machine with
  16 GB memory running macOS 10.13.4} \footnote{ProbLog version:
  2.1.0.19 and PITA available with SWI-Prolog 7.6.4}.  ProbLog
provides independent sampling along with an option to propagate
evidence. We didn't find these to be effective and ProbLog failed to
generate consistent samples in reasonable time (5 mins). PITA sampling
\cite{riguzzi2011mcintyre} on the other hand provided better
performance (see table \ref{tab:sampling-runtimes}). Due to the lack
of constraint processing, the time required per sample is higher for
PITA although the convergence behavior is similar. It should be noted
that PITA counts only consistent samples, and there can be many
attempts at generation of a consistent sample. In contrast, our
constraint processing techniques allow us to generate a consistent
sample at every attempt (for the specific examples considered). The
plots showing estimated answer probability and the variance of the
estimates for the two examples for PITA are shown in
Fig. \ref{fig:pita-problog-plots}.

\begin{table}[h]
  \centering
  \begin{tabular}{l c c c c c c}
    \hline
    Problem & OSDD gen. & LW & ProbLog & ProbLog-pe & Mcint.-rej. & Mcint.-MH\\
    \hline
    palindrome & 0.019 & 0.00017 & 188 & na & 0.41 & 7.7 \\
    birthday & 0.025 & 1.7e-5 & 19 & 26 & 0.004 & 0.003\\
  \hline
\end{tabular}
\caption{Time for OSDD computation and per consistent sample (seconds)}
\label{tab:sampling-runtimes}
\end{table}

\paragraph{PITA and ProbLog exact inference}
We evaluated the exact inference procedures of PITA and ProbLog on the same 
examples. We used a timeout of 15 minutes for both systems. 
ProbLog's 
inference does not scale beyond small problem sizes for thes two
examples.
PITA could successfully compute the
conditional probabilities for
Palindrome with $n=18$.  
PITA's inference completed for the Birthday example with population
size $2$, but ran out of memory for larger population sizes. The table
\ref{tab:birthday-palindrome-osdd-times} shows the time required to
construct the OSDDs for the birthday example and the palindrome
example. The leftmost column gives the population size/length of the
string. While the birthday problem has a single osdd, the palindrome
example requires two osdds: the one for evidence and one for
conjunction of query and evidence. It turns out that all of them
satisfy the measurability property.  However we note that the osdd for
conjunction of query and evidence for palindrome example is
intractable for large problem sizes. The evidence osdd for palindrome
on the other hand is very simple and scales well. The columns with
title ``M. prob'' show the time required to compute the probability
from the OSDDs by exploiting measurability.
\begin{table}[h]
  \centering
  \begin{tabular}{l | c c | c c c}
    \hline
    & Birthday & & & Palindrome & \\
    \hline
    Size & osdd & M. prob. &  evid. osdd & qe osdd & M. prob\\
    \hline
    6 & 0.025 & 0.005 & 0.01 & 0.043 & 0.003\\  
    8 & 0.079 & 0.008 & 0.01 & 0.208 & 0.006\\
    10 & 0.215 & 0.014 & 0.0109 & 0.7189 & 0.008\\
    12 & 0.795 & 0.024 & 0.0129 & 1.877 & 0.008\\
    14 & 5.49 & 0.035 & 0.014 & 4.94 & 0.033\\
    16 & 52.83 & 0.056 & 0.017 & 16.03 & 0.076\\
    \hline
  \end{tabular}
  \caption{Time for OSDD generation and probability computation
    (seconds)}
  \label{tab:birthday-palindrome-osdd-times}
\end{table}


\comment{overhead of up to 4 \emph{times} 
Node consistency has minimal overhead. It is $O(nd)$ in the worst case
where $n$ is the number of variables and $d$ is the size of the
largest discrete domain. The complexity of arc-consistency however
depends on the algorithm used. The best known complexity is for
arc-consistency is $O(ed^2)$ where is $e$ is the number of arcs (i.e,
binary constraints) and $d$ is the size of the largest domain. This is
achieved by the AC-4 algorithm \cite{mohr1986arc}. The AC-3 algorithm
\cite{mackworth1977consistency} has worst case complexity of $O(ed^3)$
and this was used in the implementation. The worst case complexity of
forward-checking is same as arc-consistency. In terms of actual
observed overheads, we noticed an overhead of about 4 msec for node
consistency in the Grid and PRM experiments. The sample generation
overhead in PRM, Birthday and Palindrome experiments were
multiplicative factors of 5.2, 12.5 and 4.5. The rather high overhead
for Birthday experiment can be understood from the fact, that most
samples in independent sampling encounter failure and do not evaluate
all random variables, whereas each sample of likelihood-weighted
sampler contains values of all random variables. We do not provide any
sample generation overhead for Grid and Ising model since the
independent sampler generated no consistent samples.
\comment{\begin{table}
\begin{tabular}{c c c}
\hline
Experiment & Preprocessing(msec) & Sample(factor) \\
\hline
Grid & 4msec & NA \\
PRM & 4msec & 5.2 \\
Birthday & NA &  \\
Palindrome & NA & 4.5 \\
\hline
\end{tabular}
\caption{overheads}
\label{table:overheads}
\end{table}}
}

\section{Related Work}
\label{sec:related}
Symbolic inference based on OSDDs was first proposed in
\cite{nampally2015constraint}. The present work expands on it two significant ways: Firstly, the construction of OSDDs is 
driven by tabled evaluation of transformed programs instead of by
abstract resolution. Secondly, we give an exact inference procedure for
probability computation using OSDDs which generalizes the exact inference
procedure with ground explanation graphs.

Probabilistic Constraint Logic Programming \cite{michels2013inference}
extends PLP with constraint logic programming (CLP). It allows the
specification of models with imprecise probabilities. Whereas a world
in PLP denotes a specific assignment of values to random
variables, a world in PCLP can define constraints on random
variables, rather than specific values. Lower and upper bounds are
given on the probability of a query by summing the probabilities of
worlds where query follows and worlds where query is possibly true
respectively. While the way in which ``proof constraints'' of a PCLP
query are obtained is similar to the way in which symbolic derivations are
obtained (i.e., through constraint based evaluation), the inference
techniques employed are completely different with PCLP employing
satisfiability modulo theory (SMT) solvers. \comment{The compairson of PCLP and
our constraint based evaluation is a subject of further study.}

cProbLog extends ProbLog with first-order constraints
\cite{fierens2012constraints}. This gives the ability to express
complex evidence in a succinct form. The semantics and inference are
based on ProbLog. In contrast, our work makes the underlying constraints in
a query explicit and uses the OSDDs to drive inference.

CLP($\mathcal{BN}$) \cite{costa2002clp} extends logic programming with
constraints which encode conditional probability tables. A CLP($\mathcal{BN}$) program defines a
joint distribution on the ground skolem terms. Queries
are answered by performing inference over a corresponding BN.  

There has been a significant interest in the area of lifted inference
as exemplified by the work of
\cite{poole2003first,braz2005lifted,milch2008lifted}. The main idea of
lifted inference is to treat indistinguishable \emph{instances} of random
variables as one unit and perform inference at the population level.
Lifted inference in the context of PLP has been performed by
converting the problem to parfactor representation~\cite{bellodi2014lifted} or weighted
first-order model counting~\cite{van2011lifted}. Lifted explanation graphs
\cite{nampally2016inference} are a generalization of ground
explanation graphs, which treat instances of random processes in a
symbolic fashion. In contrast, exact inference using OSDDs treats
 \emph{values} of random variables symbolically,
thereby computing probabilities without grounding the random
variables.  Consequently, the method in this paper can be used when instance-based lifting is inapplicable.  Its relationship to more recent liftable classes~\cite{Kazemi:NIPS2016} remains to be explored. 

The use of sampling methods for inference in PLPs has been
widespread.  The evidence has generally been handled by heuristics to
reduce the number of rejected samples
\cite{cussens2000stochastic,moldovan2013mcmc}.  More recently, \cite{Nitti:2016} present an algorithm that generalizes the applicability of LW samples by recognizing when valuation of a random variable will lead to query failure.  Our technique propagates constraints imposed by evidence.  With a rich constraint language and a propagation algorithm of sufficient power, the sampler
can generate consistent samples without any rejections.

Adaptive sequential rejection sampling \cite{mansinghka2009exact} is
an algorithm that adapts its proposal distributions to avoid
generating samples which are likely to be rejected. However, it
requires a decomposition of the target distribution, which may not be
available in PLPs. Further, in our work the distribution from which
samples are generated is not adapted. It is an interesting direction
of research to combine adaptivity with the proposed sampling
algorithm.

\section{Conclusion}
\label{sec:conclusion}
In this work we  introduced OSDDs as an alternative data structure for PLP.  OSDDs enable efficient inference over programs whose random variables range over large finite domains.    We also showed the effectiveness of using OSDDs for likelihood weighted sampling.

OSDDs may provide asymptotic improvements for inference over many classes of first-order probabilistic graphical models.  An example of such models is the Logical hidden Markov model (LOHMM) which lifts the representational structure of hidden Markov models (HMMs) to a first-order domain \cite{Kersting06logicalhidden}.  LOHMMs have proved to be useful for applications in computational biology and sequential behavior modeling.  LOHMMs encode first-order relations using \emph{abstract transitions} of the form $p : \mathtt{H} \xleftarrow[]{\mathtt{O}} \mathtt{B}$ where $p \in [0, 1]$.  \comment{An abstract transition can be read as ``starting in state $\mathtt{B}$, transition to state $\mathtt{H}$ and emit observation $\mathtt{O}$ with probability $p$".  The set of abstract transitions $p_i : \mathtt{H}_i \xleftarrow[]{\mathtt{O}_i} \mathtt{B}$ which share a body term $\mathtt{B}$ must satisfy that $\sum_i p_i = 1$.  Thus abstract transitions define a distribution $\Delta$ over the possible transitions from state $\mathtt{B}$.
}
Any of $\mathtt{H}, \mathtt{B}, \mathtt{O}$ may be partially ground, and there may be logical variables which are shared between any of the atoms in an abstract transition.  Abstract explanations that are obtained by inference over such models that avoids grounding of variables whenever possible can be naturally captured by OSDDs. 

\comment{
These logical variables are tied by unification, requiring a standard grounding order to be introduced over $\mathtt{H}, \mathtt{B}, \mathtt{O}$.

As an example, the observation sequence may model the purchasing habits of some user and be of the form $\mathtt{O} = buys(X), sells(X'), ..., saved(Y), buys(Y)$.  Observations may have logical variables which unify in the processing of some abstract transition, say $0.35 : sells(X', john) \xleftarrow[]{buys(X)} buys(X, john)$.  The sequence of observation variables introduces a constraint sequence, $X \neq X', ..., Y = Y$.  We believe that OSDDs may provide an elegant framework for probabilistic inference over LOHMMs due to these constraint sequences.  
}

\bibliography{iclp2018}

\newpage
\appendix
\section{Proofs}
\begin{Pro1}[Closure properties]
	OSDDs are closed under conjunction and disjunction
	operations.
	\begin{proof}
		Let $\psi = (s, k, Y)[\gamma_i : \psi_i]$ and
		$\psi' = (s', k', Y')[\gamma'_j : \psi'_j]$ be two OSDDs.
		
		Let $\oplus$ denote either $\land$ or $\lor$, then by the
		definition of $\psi \oplus \psi'$ ordering is preserved.
		Depending on the ordering of the OSDDs, $\psi \oplus \psi'$ has
		three cases. If $(s, k) \prec (s', k')$ (resp.
		$(s', k') \prec (s, k)$ then $\psi \oplus \psi'$ is constructed by
		leaving the root and edge lables intact at $(s, k, Y)$
		(resp. $(s', k', Y'))$. In this case urgency, mutual exclusion,
		and completeness are all preserved since $\psi$ (resp. $\psi'$) is
		an OSDD and the root and its edge labels are unchanged.
		
		If $(s, k) = (s', k')$ urgency is preserved since
		$\forall i \forall j$ $\gamma_i \land \gamma'_j$ are the
		constructed edges of $\psi \oplus \psi'$ and individually these
		$\gamma_i$ and $\gamma'_j$ satisfied urgency. If we take two
		distinct edge constraints $\gamma_i \land \gamma'_j$ and
		$\gamma_k \land \gamma_l$ it is the case that
		$\sem{\gamma_i \land \gamma'_j \land \gamma_k \land \gamma'_l} =
		\emptyset$ since either $i \neq k$ or $j \neq l$ and both
		$\sem{\gamma_i \land \gamma_k} = \emptyset$ and
		$\sem{ \gamma'_j \land \gamma'_l} = \emptyset$. Let $\sigma$ be
		the grounding substitution of
		$\cup_{i,j} Vars(\gamma_i \land \gamma'_j) \setminus \{Y\}$ that
		is compatible with constraint formula labeling the path to
		the node $(s,k,Y)$. To prove completeness, we note that
		$\cup_j \sem{\gamma_i\sigma \land \gamma'_j\sigma}_Y =
		\sem{\gamma_i\sigma}_Y$. Therefore,
		$\cup_{i,j} \sem{(\gamma_i \land \gamma'_j)\sigma} = type(Y)$. \hfill
	\end{proof}
\end{Pro1}

\begin{Pro1}
  Let $\psi = (s, k, Y)[\gamma_i : \psi_i]$ and
  $\psi' = (s', k', Y')[\gamma'_j : \psi'_j]$ be two OSDDs, then
  \begin{gather*}
    \mathcal{G}(\psi \oplus \psi') = \mathcal{G}(\psi) \oplus \mathcal{G}(\psi').
  \end{gather*}

  \begin{proof}
    When $(s, k) \prec (s', k')$, then
    $\mathcal{G}(\psi) \oplus \mathcal{G}(\psi') =
    (s,k,Y)[\alpha_r:\mathcal{G}(\psi_r[\alpha_r/Y]) \oplus
    \mathcal{G}(\psi')]$. But $\mathcal{G}(\psi \oplus \psi') =
    \mathcal{G}((s,k,Y)[\gamma_i: \psi_i \oplus \psi']) =
    (s,k,Y)[\alpha_r: \mathcal{G}(\psi_r \oplus \psi'
    [\alpha_r/Y])]$. 

    Thus, we consider the case where $(s, k) = (s', k')$.  Both
    ground explanation graphs have the same root, therefore the ground
    explanations in $\mathcal{G}(\psi) \oplus \mathcal{G}(\psi')$ are
    obtained by combining subtrees connected which have the same edge
    label. Given grounding substitution $\sigma$ on
    $\cup_{i,j}Vars(\gamma_i \land \gamma'_j)\setminus\{Y\}$ that is
    compatible with the constraint formula labeling the path from root
    to the node under consideration, if some value
    $\alpha \in type(Y)$ is such that it satisfies $\gamma_i\sigma$
    and $\gamma'_j\sigma$ for specific $i,j$, then in
    $\psi \oplus \psi'$,
    $\alpha \in \sem{(\gamma_i \land \gamma_j)\sigma}_Y$, therefore
    the same subtrees are combined.  
  \end{proof}
    
\end{Pro1}

\begin{Pro1}[Condition for Measurability]
	A satisfiable constraint formula is measurable w.r.t all of its
	variables if and only if it saturated.
	\begin{proof}
		First we prove that saturation is a sufficient condition for
		measurability.
		
		The proof is by induction on the number of variables in
		$\gamma$. When $|Vars(\gamma)|=1$ the proposition holds since the
		only satisfiable constraint formulas with a single variable are
		$\{X=c\}$ for some $c \in Dom(X)$ or formulas of the form
		$\{X \neq c_1, X\neq c_2,\ldots, X \neq c_m\}$ for some distinct
		set of values $\{c_1,\ldots,c_m\} \subset Dom(X)$. Clearly the
		formulas are measurable w.r.t $X$.
		
		Assume that the proposition holds for saturated constraint formulas with $n$
		variables. Now consider a satisfiable
		constraint formula $\gamma$ with $n+1$ variables which is saturated. Let $X \in Vars(\gamma)$. Consider the graph
		obtained by removing $X$ and all edges incident on $X$ from the
		constraint graph of $\gamma$. It represents a saturated constraint formula
		$\gamma'$ with $n$ variables. This is
		because for any three variables $A, B, C$ distinct from $X$, if $A
		= B, B = C$ then $A, C$ are connected by an ``$=$" edge. Similarly, if
		$A = B, B \neq C$, then $A, C$ are connected by an ``$\neq$" edge. Further
		for any variable $A$ other than $X$, if $Z$ is the set of
		variables connected to $A$ by ``$\neq$" edges, then there exists edges
		between each pair of these nodes. This is due to the definition of
		saturation which is satisfied by $\gamma$.
		
		But, by inductive hypothesis $\gamma'$ is measurable w.r.t each of
		its variables. Now consider computing the measure of $X$ in
		$\gamma$.  If $X$ is connected to any node $Y$ with an ``$=$" edge,
		then measure of $X$ is 1. If $X$ is not connected to any
		node with an ``$=$" edge, then it is either disconnected from other
		nodes or connected to them by only ``$\neq$" edges. In either case $m_X$
		is computed by subtracting the number of nodes connected to $X$ by
		``$\neq$"  edges from the domain.
		
		To prove that saturation is a necessary condition we use proof by
		contradiction.  Assume there exists a measurable constraint
		formula $\gamma$ which is not saturated. Then there
		exists a variable $X \in Vars(\gamma)$ and a set $\mathcal{Z}$ which is the
		set of nodes connected to the node for $X$ by ``$\neq$" edge and for
		some pair of elements $A, B \in \mathcal{Z}$, there is no edge between
		them. Since we take closure of ``$=$"  edges, we can assume that
		$\gamma \not \models A=B$. So there must exist two substitutions
		$\sigma, \sigma'$ where $A=B$ and $A\neq B$ respectively. The
		number of solutions of $X$ under these two substitutions is
		clearly different, which is a contradiction.
	\end{proof}
\end{Pro1}

\end{document}